# Revealing the Microscopic Mechanism of Elementary Vortex Pinning in Superconductors

C. Chen[1,10†], Y. Liu[4,8†], Y. Chen[3†], Y. N. Hu[1], T. Z. Zhang[1,11], D. Li[4,8], X. Wang[1], C. X. Wang[1], Z.Y.W. Lu[4,8], Y. H. Zhang[4,8], Q. L. Zhang[1], X. L. Dong[4,8,9], R. Wang[3,6]*, D. L. Feng[2,1,6,7]*, T. Zhang[1,5,6,7]*

[1]Department of Physics, State Key Laboratory of Surface Physics and Advanced Material Laboratory, Fudan University; Shanghai 200438, China
[2]New Cornerstone Laboratory, National Synchrotron Radiation Laboratory and School of Nuclear Science and Technology, University of Science and Technology of China; Hefei 230027, China
[3]National Laboratory of Solid State Microstructures and Department of Physics, Nanjing University; Nanjing 210093, China
[4]Beijing National Laboratory for Condensed Matter Physics, Institute of Physics, Chinese Academy of Sciences; Beijing 100190, China
[5]Hefei National Laboratory, Hefei 230088, China
[6]Collaborative Innovation Center for Advanced Microstructures; Nanjing 210093, China
[7]Shanghai Research Center for Quantum Sciences; Shanghai 201315, China
[8]School of Physical Sciences, University of Chinese Academy of Sciences; Beijing 100049, China
[9]Songshan Lake Materials Laboratory; Dongguan, Guangdong 523808, China
[10]Zhejiang Institute of Photoelectronics & Zhejiang Institute for Advanced Light Source, Zhejiang Normal University; Jinhua 321004, China.
[11]Interdisciplinary Materials Research Center, School of Materials Science and Engineering, Tongji University; Shanghai 201804, China

† These authors contributed equally.
*Corresponding authors: rwang89@nju.edu.cn, dlfeng@ustc.edu.cn, tzhang18@fudan.edu.cn

**Vortex pinning is a crucial factor that determines the critical current of practical superconductors and enables their diverse applications. However, the underlying mechanism of vortex pinning has long been elusive, lacking a clear microscopic explanation. Here using high-resolution scanning tunneling microscopy, we studied single vortex pinning induced by point defect in layered FeSe-based superconductors. We found the defect-vortex interaction drives low-energy vortex bound states away from $E_F$, creating a "mini" gap that effectively lowers the system energy and enhances pinning. By measuring the local density-of-states, we directly obtained the elementary pinning energy and estimated the pinning force via the spatial gradient of pinning energy. The results are consistent with bulk critical current measurement. Furthermore, we show that a general microscopic quantum model incorporating defect-vortex interaction can naturally capture our observation. It suggests that the local pairing near pinned vortex core is actually enhanced compared to unpinned vortex, which is beyond the traditional understanding that non-superconducting regions pin vortices. Our study thus unveils a general microscopic mechanism of vortex pinning in superconductors, and provides insights for enhancing the critical current of practical superconductors.**

## I. INTRODUCTION

The ability to carry electric current without dissipation is a defining property of superconductivity. However, in practical superconductors (mostly type II superconductors), external current generates Lorentz force on quantized magnetic flux and dissipation occurs when the associated vortices move [Fig. 1(a)]. Fortunately, it has been found that defects or disorders can prevent the motion of vortices, known as flux/vortex pinning effect. The critical current density ($J_C$) actually depends on vortex pinning strength rather than current induced de-pairing [1, 2]. Therefore, understanding the mechanism of how a single vortex is pinned by defect, namely the elementary vortex pinning, is of fundamental importance for technological use of superconductors [3-5]. It also lays the foundation of vortex dynamics which determines the full electromagnetic response of superconductors.

In the general description, a vortex has a non-superconducting core with a size of $2\xi$ [$\xi$ is the coherence length, as shown in Fig. 1(b)]. Traditional understanding of vortex pinning is based on Ginzburg-Landau (G-L) theory [1,2], which consider the pinning centers as non-superconducting regions and the vortex cores attached to them will save the condensation energy cost. However, such phenomenological treatment is difficult to describe pinning centers much smaller than $\xi$, such as point-like defects (e.g., impurity atom or vacancy) which represent the elementary form of defects in practical materials. Some theoretical works have incorporated an impurity term into G-L free-energy to address this issue [6-8]. In fact, the microscopic description of vortex beyond the G-L theory has already been given by Caroli, de Gennes and Matricon (CdGM) [9], which predicted localized bound states with discrete energies ($E = \mu\Delta^2/E_F$, $\mu= \pm1/2, \pm3/2…$) in the vortex core [Fig. 1(b)]. Meanwhile, impurity induced effect in superconductors was also extensively studied [10]. One may expect the local interaction between CdGM states and defect shall play an important role in vortex pinning. The theoretical work by Q. Han, L.Y. Zhang and Z. D. Wang [11] has suggested such scenario, and a recent work has predicted shifted CdGM state due to impurity-vortex interaction [12]. However, for a long time the study of vortex pinning has been limited to indirect transport or force measurements [1-4,13-17]. Due to the difficulty of direct investigation on the electron structure/energy of a single vortex and its pinning center (defect), the understanding of vortex pinning is still phenomenological so far.

Scanning tunnelling microscopy/spectroscopy (STM/STS) with atomic resolution is a powerful tool to study the microscopic structure of a single vortex [18]. Particularly, recent high-resolution tunneling spectrum has been able to identify discrete CdGM vortex states in various FeSe-based superconductors [19-23], as well as atomic defects induced states [24-26]. This enables direct measurement of vortex–impurity interaction in the atomic scale. In this work, we performed detailed STM study on the vortex pinning in (Li, Fe)OHFeSe and single-layer FeSe/SrTiO$_3$. We found that when a single vortex is pinned by a point defect in FeSe layer, the low-energy CdGM states are "pushed" away from $E_F$, which lowers the formation energy of vortices and causes pinning. It indicates an enhanced local pairing at the pinned vortex core (with respect to that in unpinned vortex), which is beyond the traditional understanding that it is the non-superconducting region that pins the vortex. Such novel pinning mechanism is well captured by our microscopic quantum model describing the vortex-impurity interaction, which can be applied to general superconductors. Furthermore, via measuring the local density-of-state (DOS) distribution over the pinned/unpinned vortex cores, we are able to obtain the elementary pinning energy for the first time. We further tuned the defect-vortex distance via the repulsive vortex-vortex interaction, then the pinning force is estimated as the spatial gradient of pinning energy. The obtained value (~$2\times10^{-4}$ N/nm) aligns with the bulk critical current measurement. Our work thus sets up the microscopic mechanism of vortex pinning induced by point defect.

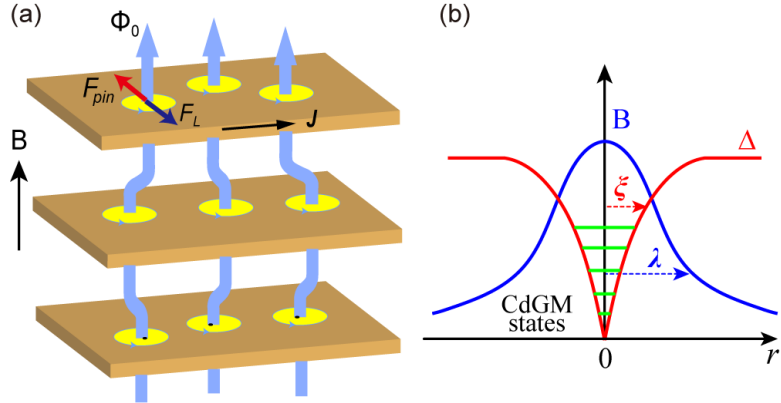

**FIG. 1.** Schematics of vortex structure and vortex pinning. (a) Illustration of vortex pinning in layered superconductor. The magnetic flux lines are pinned by individual point defects in each superconducting layer. (b) The structure of a single vortex core.

## II. EXPERIMENTAL RESULTS
### A. The local tunneling spectra of pinned/unpinned vortex cores

The STM experiment was conducted using a dilution refrigerator STM (Unisoku) at the base temperature of 20 mK ($T_{eff}$ = 160 mK), or at 4.2K when specified. The samples studied here are optimally doped (Li, Fe)OHFeSe ($T_C$ = 42K) single crystalline film [27,28] and single-layer FeSe/SrTiO$_3$ film. Details of the sample preparation are described in Method section. As sketched in Fig. 2(a), (Li, Fe)OHFeSe crystal consists of alternatively stacked FeSe and (Li, Fe)OH layers. It is important to note that recent transport studies of (Li, Fe)OHFeSe [29] evidenced pancake-like vortices in FeSe layers, as depicted in Fig. 1(a). The interlayer vortex coupling is weak due to (Li, Fe)OH intercalation.

Figure 2(b) shows a topographic image of the FeSe-terminated surface of cleaved (Li, Fe)OHFeSe film. The most obvious and commonly observed defects in FeSe layer are the "dumbbell"-shaped defects indicated by green arrows [see Fig. 2(c) for zoomed-in image]. They are mostly Fe vacancies formed during sample synthesis [28]. On the defect-free region, a full superconducting gap with two pairs of coherence peaks at ±8 mV and ±14 mV is observed [Fig. 2(d)]. The flat gap bottom has a half-width of 5.5 meV, corresponding to the minimum gap value. The dumbbell defects can induce multiple pronounced impurity states inside of superconducting gap [Fig. 2(d)], which originated from their strong and anisotropic scattering potential, as discussed in ref. 26.

Upon applying an out-of-plane magnetic field of B=11T, vortex cores are visualized in the zero-bias dI/dV map displayed in Fig. 2(e). Notably, more than 50% of vortex cores are located at the positions of dumbbell defects (indicated by green arrows), which makes the vortex lattice highly distorted. Particularly, the spacings between some of these vortex cores are considerably smaller than the averaged vortex-vortex spacing at B =11T (14.7nm), as indicated in Fig. 2(e). This implies the dumbbell defects are strong pinning centers of vortex. Additionally, vortices are also found in defect-free regions and we refer them as "free" vortices (indicated by yellow arrows, see also Fig. S2 for additional vortex maps). Interestingly, all the pinned vortices display a "dark spot" near their centers (defect site). As shown below, this feature results from suppressed low-energy CdGM state at the defect site, which is a key indicator of vortex pinning. We note that besides the dumbbell defects, there also exist Se vacancies on the surface [26], but they do not show significant pinning effect on vortex cores (see Fig. S1 and Part I-1 of supplementary materials for more details).

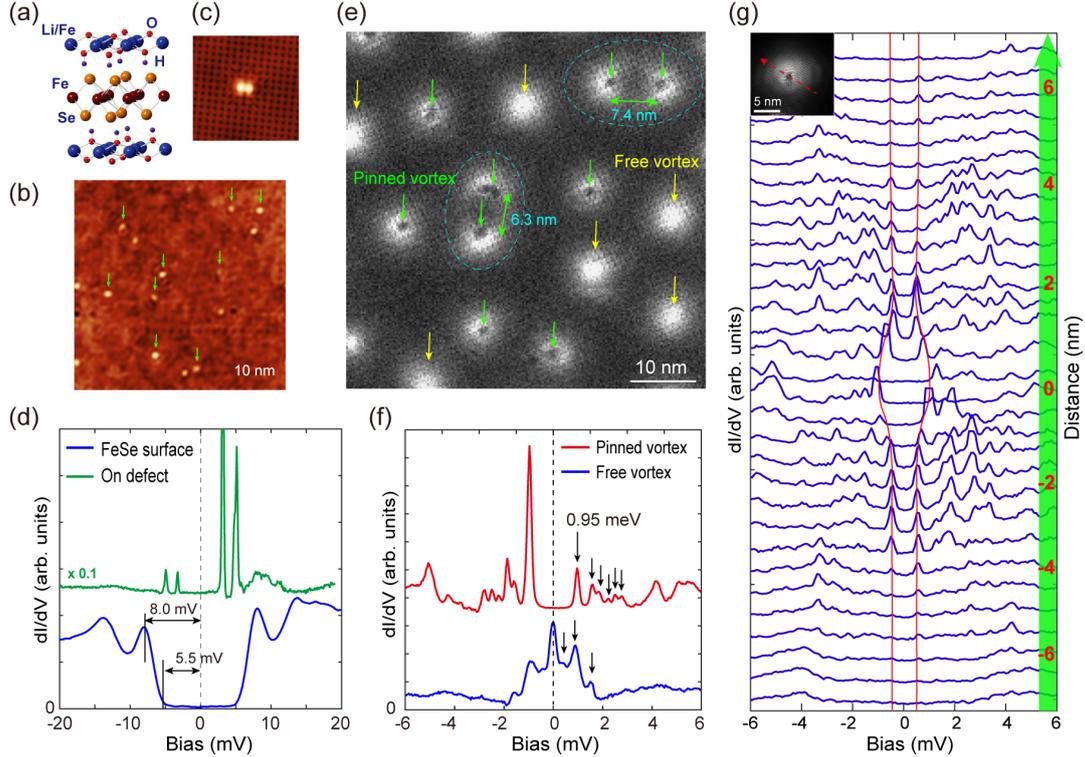

**FIG. 2.** (a) Crystal structure of (Li, Fe)OHFeSe, composed of FeSe layers and (Li,Fe)OH layers. (b) Topographic image of the FeSe surface of (Li,Fe)OHFeSe ($V_b$ = 50mV, $I$ = 60pA). (c) Atomically resolved image of a dumbbell-defect at Fe cite. (d) Typical dI/dV spectra on dumbbell-defect and defect-free region at B=0T ($V_b$ = 20mV, $I$ = 60pA). (e) Zero-bias dI/dV map at B=11T ($V_b$ = 40mV, $I$ = 40pA, T=4.2K), taken at the same region of panel (b). Green arrows in (b) and (e) indicate the positions of dumbbell defects and the pinned vortices on these defects. Yellow arrows in (e) indicate the free (unpinned) vortices. (f) Typical dI/dV spectra taken at the center of pinned and free vortex cores ($V_b$ = 17mV, $I$ = 60pA). Arrows indicate the in-gap states. (g) A series of dI/dV spectra taken along the red arrow across the pinned vortex core shown in the inset image ($V_b$ = 10mV, $I$ = 60pA). The red curves track the position of the lowest CdGM states.

Fig. 2(f) shows high-energy resolved dI/dV spectra taken at the center of pinned and free vortex cores. For free vortices, a zero-bias conductance peak (ZBCP) with a series of CdGM states around $E_F$ are observed. The ZBCP was shown to have characteristics of Majorana zero mode [30,31]. Remarkably, in the pinned vortex the ZBCP and nearby CdGM state are absent, resulting in a "mini gap" between ±0.95 mV while a large number of discrete peaks appear outside of this gap. Fig. 2(g) shows a series of dI/dV spectra taken across a pinned vortex core. It's seen that the mini gap has the largest size at the center (defect site), but rapidly decreases to a constant value at ~ 1.0 nm away from the pinning center. This behavior gives rise to the "dark spot" in the zero-bias dI/dV map, and evidences that the local vortex-impurity interaction drives the low-energy CdGM state away from $E_F$, since the distribution of impurity state is very localized (with a scale shorter than coherence length) [24-26]. The discrete peaks outside of mini gap are likely from the hybridization between CdGM state and impurity state.

A direct consequence of a mini gap opening in vortex core is that the formation energy (cost of condensation energy) of a pinned vortex is reduced with comparing to that of a free vortex. This is essentially why the defects can pin vortices here. The absence of ZBCP could be due to the dumbbell defects (Fe vacancies) are strong magnetic impurities that may locally disrupt the topological band structure of (Li, Fe)OHFeSe [21]. Meanwhile, the dumbbell defects observed in topographic image should only pin the "pancake" vortex in the topmost FeSe layer [Fig. 1(a)], whether the vortices are affected by underneath defect is unclear but expected to be weak. We then further examined another FeSe-based superconductor, the single-layer FeSe/SrTiO$_3$ which has only one FeSe layer and exhibits conventional CdGM states in the vortex cores [22, 32].

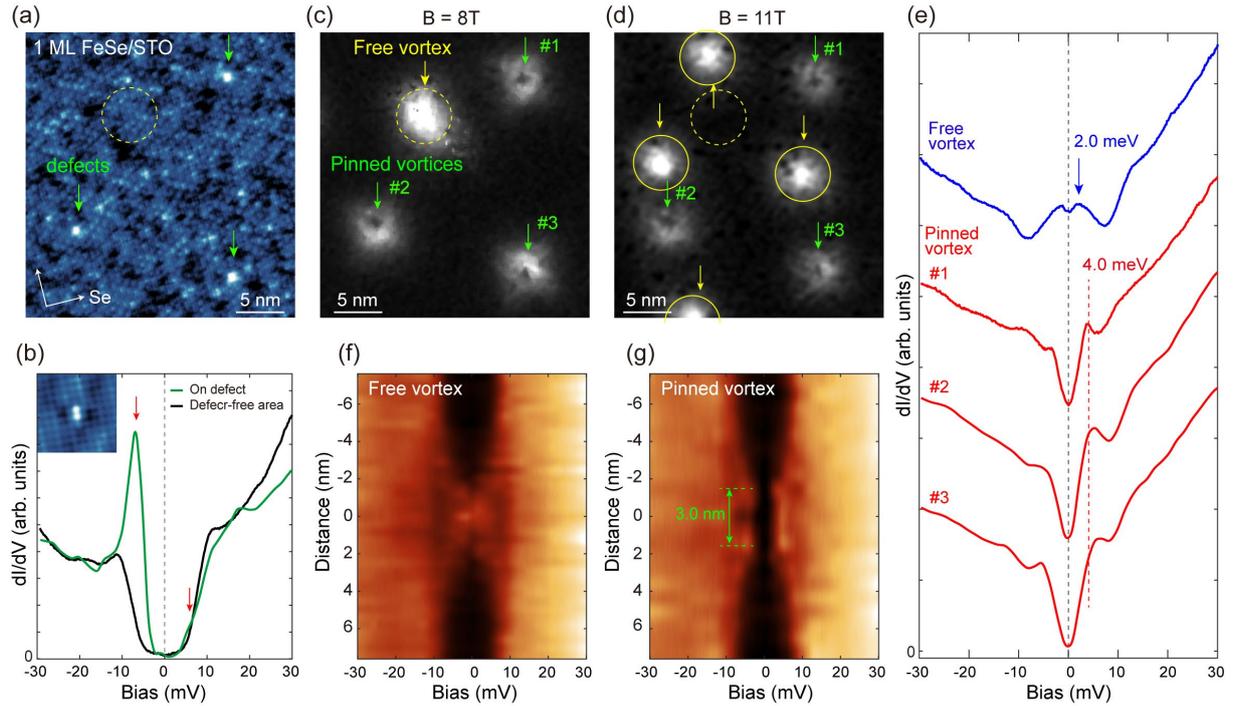

**FIG. 3. STM investigation of vortex pinning in 1ML FeSe/SrTiO$_3$ (001).** (a) Topographic image of 1ML FeSe/SrTiO$_3$, with green arrows indicating three "dumbbell-like" defects ($V_b$ = 28.5mV, $I$ = 100pA). (b) The dI/dV spectra taken at the dumbbell defect (inset image) and defect-free area at B=0T. Red arrows indicate the impurity states. (c,d) Zero-bias dI/dV map taken at the same region of panel (a) under B = 8T and B=11T, respectively. Green arrows indicate pinned vortices and yellow arrows indicate free vortices (setpoint: $V_b$ = 28.5mV, $I$ = 100pA, T = 4.2K). (e) dI/dV spectra measured at the center of free vortex, pinned vortices ($V_b$ = 30mV, $I$ = 80pA, T = 4.2K). (f,g) Color plots of the dI/dV spectra taken across the centers of free vortex and pinned vortex, respectively ($V_b$ = 30mV, $I$ = 80pA, T = 4.2K).

Fig. 3(a) shows a topographic image of single-layer FeSe/SrTiO$_3$. One can also observe dumbbell-like defects locate at the Fe sites, as indicted by green arrows [a zoomed-in image is shown in the inset of Fig. 3(b)]. These defects could be either Fe vacancies or impurity atoms formed during MBE growth. They also produce pronounced impurity states within the superconducting gap [Fig. 3(b)]. Fig. 3(c) shows the zero-bias dI/dV map taken at the same region of Fig. 3(a) under B = 8T. Clearly, there are three vortex cores located at the dumbbell defect sites (labelled as #1-3), each displaying a "dark spot" at its center. Additionally, there is one free vortex located at defect-free region (marked by dashed circle). We then increased the magnet field to 11T and acquired another vortex map at the same region [Fig. 3(d)]. The free vortex at B=8T has changed its position, and more vortices emerge at defect-free areas. Meanwhile the vortices #1-3 remain at the defect sites, further indicating they are pinned vortices.

In Fig. 3(e), we plot the dI/dV spectra taken at the pinned vortices #1-3 and a free vortex. Although these spectra are acquired at T = 4.2K, the energy shifts of lowest CdGM states of pinned vortices are clearly visible (the shift is approximately 2.0 mV). Figs. 3(f) and 3(g) display color plots of the dI/dV spectra taken across a free and a pinned vortex, respectively. Similar to that observed in (Li, Fe)OHFeSe, the "mini gap" is also localized around defect site within a region approximately ±1.5 nm. Therefore, the vortex pinning behaviour in single-layer FeSe/SrTiO$_3$ is similar to that of (Li, Fe)OHFeSe.

## B. Extracting elementary pinning energy/force from tunneling spectra

A key quantity reflects the vortex pinning strength is the elementary pinning energy ($U_{pin}$), which is the energy difference between a pinned vortex and a free vortex. For point-like defects, the dominant part of pinning energy is expected to originate from condensation energy [1,2,11] (the magnetic penetration length of (Li,Fe)OHFeSe is ~160 nm [36], which is much larger than the coherence length (~2 nm) and the scale of point defect, thus the influence of point defect to surrounding magnetic field and supercurrent distribution would be negligible). It is known that the tunneling conductance (dI/dV) is proportional to the local DOS (LDOS), thus the total energy of a vortex can be obtained by integrating calibrated dI/dV spectra over energy ($E$) and the area ($S$) covers the vortex core. Then the $U_{pin}$ can be quantitively calculated via:

$$U_{pin} = \int \int_{E_{LB}}^{E_F} N(0) \left[ \left( \frac{dI}{dV} \right)_{pin} - \left( \frac{dI}{dV} \right)_{free} \right] E dE dS$$

Here the d$I$/d$V$ spectra is normalized by its value at the energy of $E_{LB}$, which is below the superconductivity gap edge and is the lower bound of integration ($E_{LB}$ = -17 meV is used). $N(0)$ is the absolute value of normal state DOS (per area) near $E_F$. The Fermi surfaces of (Li, Fe)OHFeSe of FeSe/SrTiO$_3$ are both composed of two electron pockets at M point [32-35], thus $N(0)$ can be obtained from the band dispersion acquired by quasi-particle interference (QPI) measurement (see Part I-2 of supplementary materials). The calculated value of $U_{pin}$ for (Li,Fe)OHFeSe and FeSe/SrTiO$_3$, are -1.8 meV and -2.3 meV, respectively. To our knowledge, this is the first direct measurement of elementary pinning energy. If taking the spatial region where the mini-gap opens as an effective "pinning radius" ($r_p$), which is ≈1.0 nm for (Li, Fe)OHFeSe and ≈1.5 nm for 1ML FeSe/SrTiO$_3$, the pinning force can be estimated by $\bar{f}_p = U_{pin}/r_p$ ≈ 1.5-1.8×10$^{-13}$N for a single pancake vortex in FeSe layer.

A more precise way to measure $f_p$ is through the spatial gradient of $U_{pin}$. When a pinned vortex core moves away from pinning site, $|U_{pin}|$ shall decrease and yield $f_p = -\partial U_{pin}/\partial d$ [Fig. 4(a)]. Here we managed to "push" a pinned vortex of (Li, Fe)OHFeSe via repulsive vortex-vortex interaction. Fig. 4(b) shows a topographic image with a few dumbbell defects (marked by numbers). Under a vertical field of B = 6T, two pinned vortices show up at defect 1 and defect cluster 3 [Fig. 4(c)]. When the field increased to 7T [Fig. 4(d)], due to the rearrangement of vortex lattice, the vortex on defect cluster 3 "jumped" to defect cluster 2. Since the distance between vortex 1 and vortex 2 (10 nm) is significantly shorter than the averaged inter-vortex distance at 7T (18 nm), the repulsive interaction makes vortex 1 shift slightly with respected to its position at B = 6T. To see this spatial shift more clearly, Figs. 4(e)-(g) show the zoomed-in image and corresponding dI/dV maps (at E = 5 meV) around defect 1 under B = 6T, 7T, respectively. The center of the pinned vortex is determined by the ring like high-energy CdGM state distribution in Figs. 4(f)-(g). It's seen that at B = 6T there is already a small displacement between vortex center and the defect (d ≈ 0.8nm) and such displacement increased to 1.79 nm at B = 7T.

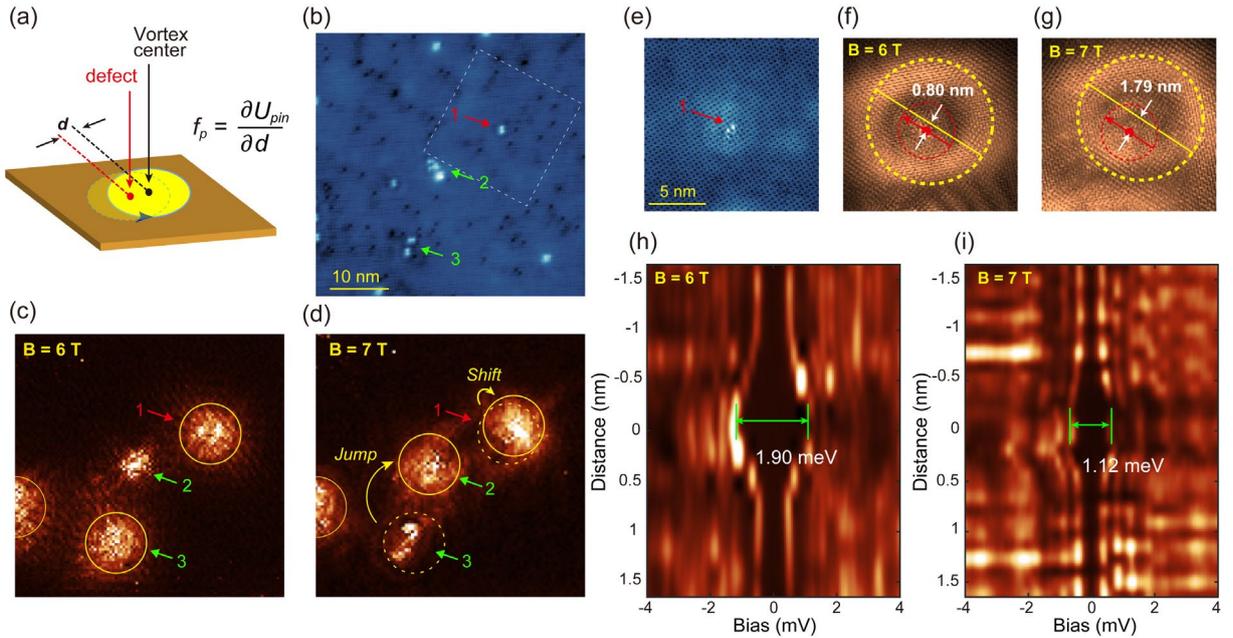

**FIG. 4. Estimation of elementary pinning force.** (a) Illustration of a pinned vortex with a distance of *d* away from pinning site. The pinning force is given by $f_p = \partial U_{pin}/\partial d$. (b) Topographic image of an FeSe surface of (Li,Fe)OHFeSe, with red and green arrows indicating typical surface impurities (V$_b$ = 100mV, *I* = 10pA). (c,d) Zero-bias dI/dV maps under magnetic fields of 6T (panel c) and 7T (panel d) in the same region of panel (b). The spatial distribution of vortices varies under different magnetic fields (V$_b$ = 30mV, *I* = 60pA, T$_{eff}$ = 160 mK). (e) Topographic image taken in the dashed box in panel (b) (V$_b$ = 10mV, *I* = 60pA). (f,g), dI/dV map taken at E = 5meV, under 6T and 7T in the same region of panel (e) (V$_b$ = 10mV, *I* = 60pA). Yellow dashed circles track the distribution of high energy CdGM state, and red dot is the position of impurity. (h,i) Color plots of a series of spectra taken along the positions indicated by the red arrows in panels (f) and (g), respectively (V$_b$ = 10mV, *I* = 60pA, T$_{eff}$ = 160 mK).

Figures 4(h)-(i) show the tunneling spectra across defect 1 at B = 6T and 7T, respectively. Notably, the defect induced mini gap at B = 7T is smaller than that at B = 6T (reduced from 1.9 meV to 1.1 meV), which indicates a lowered pinning energy at B = 7T. We quantitively calculate the $U_{pin}$ difference via dI/dV maps taken over the whole vortex (see Part I-3 of supplementary materials). The averaged pinning force in the range of 0.8nm < d <1.79 nm is then obtained by: $\bar{f}_p = |\frac{U_{pin}(6T) - U_{pin}(7T)}{d_{(7T)} - d_{(6T)}}|$, which is 2.3×10⁻¹³ N. Considering this pinning force is applied to a single "pancake" vortex in a FeSe layer [Fig. 1(a)], the elementary pinning force per unit length for bulk (Li, Fe)OHFeSe (with a c-axis constant of 0.93 nm) is $f'_p \approx 2.4 \times 10^{-4}$ N/m.

### C. Transport measurements of $J_C$ and the pinning force

Assuming all the vortices are pinned by the same type of dumbbell defects at low fields, the bulk $J_C$ should be determined by the maximum elementary pinning force, via $\boldsymbol{f'_p = J_C \times \Phi_0}$. We conducted $J_C$ measurement on the (Li, Fe)OHFeSe sample to evaluate $f'_p$. Fig. 5(a) shows the magnetic field dependence of $J_C$ at various temperatures. At T = 4K, the variation of $J_C$ is slow at B < 5T, indicating a sufficient number of pinning sites below this vortex density. This is consistent with the vortex map at B =11T [Fig. 2(e)], where over half of vortices are pinned (thus for B<5T, all the vortices in this area will be pinned). Fig. 5(b) shows the pinning force density ($\boldsymbol{F_p = J_C \times B}$) as function of B. A linear fit to the low field region (B < 2T) yields $f'_p = \frac{\partial F_p}{\partial B} \Phi_0 \approx 0.8 \times 10^{-4}$ N/m. Therefore, the elementary pinning force obtained from our tunneling measurement, 2.4×10⁻⁴ N/m, already gives a reasonable estimate on $f'_p$. For comparison, traditional strong pinning theory [2] gives $f'_p = 0.45\pi\xi\mu_0 H_c^2 = 0.45 \frac{\Phi_0^2}{8\pi\xi\lambda^2\mu_0}$, which yields a $f'_p$ of 8.0×10⁻⁴ N/m (taken $\xi$ = 2.6 nm and $\lambda$ = 160 nm for (Li, Fe)OHFeSe film [36]). The relatively large value of the pinning force we obtained compared to transport measurement could be due to that there still exists interlayer coupling between the pancake vortices in (Li, Fe)OHFeSe [Fig. 1(a)]. Since the defects are randomly distributed, the pinning forces applied on pancake vortices of adjacent FeSe layers may be partially canceled if the vortex cores are not vertically aligned (flux line rigidity induced summation problem [1,2]).

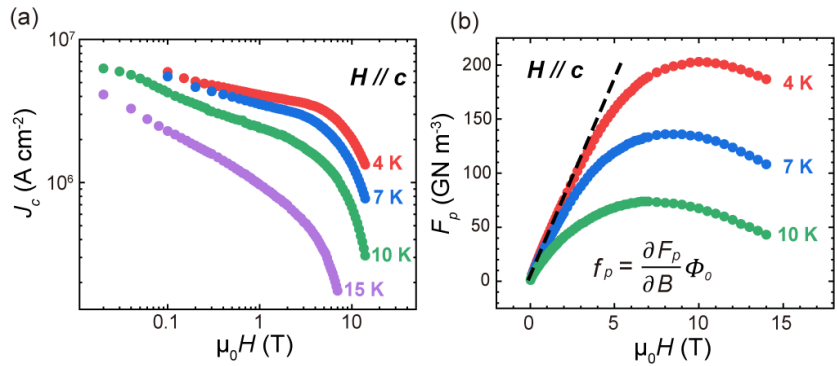

**FIG. 5.** Transport measurement of (Li,Fe)OHFeSe. (a) Magnetic field dependence of $J_c$ at different temperatures. (b) The corresponding pinning force density $F_p$ derived from Jc in panel (a). The black dashed line in (b) is a linear fit to the low field region.

### III. MICROSCOPIC MODELING OF VORTEX PINNING

So far, we have directly detected vortex-defect interaction which shifts CdGM states and show that it is responsible for vortex pinning. The mini-gap opening at pinned vortex cores [Figs. 2(g) and 3(e)] is in contrast with the phenomenological description that a non-superconducting region pins the vortex [1, 2]. To elucidate the underlying mechanism, we carried out microscopic model calculations. The system consists of a conventional superconductor with a vortex and a defect (both located at $r = 0$). The superconducting state is described by the continuum Hamiltonian $H_{SC} = \sum_\sigma \int dr[(-\hbar^2\nabla^2/2m - \mu)c_{r\sigma}^\dagger c_{r\sigma} + \Delta(r)c_{r\sigma}^\dagger c_{r\bar\sigma}^\dagger + \Delta^*(r)c_{r\bar\sigma}c_{r\sigma}]$. The local pairing potential is given by $\Delta(r) = \Delta(r)e^{iv\theta}$, where $\theta$ is the angle of $r$, and $v = 1$ characterizes the SC phase winding. The local gap function takes the form $\Delta(r) = \Delta_0 r/\sqrt{r^2 + \xi^2}$ [37]. $\Delta_0$ is the pairing potential far away from the vortex core and $\xi$ is the local coherent length. We describe the defect by a quantum impurity, $H_{imp} = \sum_\sigma \varepsilon_d d_\sigma^\dagger d_\sigma$, where $\varepsilon_d$ is the impurity level and $d_\sigma^\dagger(d_\sigma)$ is the creation (annihilation) operator of the local impurity state. The hybridization between the quantum impurity and the electrons is given by $H_{hyb} = \sum_\sigma \int dr[V_0(r)d_\sigma^\dagger c_{r\sigma} + h.c.]$, where the coupling decays with the distance, $V_0(r) = V_0 e^{-(r/r_0)^2}/\sqrt{\pi}r_0$ with $r_0$ being the decay length [12]. We note that, compared to ref. 12 which deals with a classical impurity, our quantum impurity model involves the charge fluctuations. As shown below, this generates a distinct evolution behavior of the CdGM states consistent with our experiments. The above model $H_{SC} + H_{imp} + H_{hyb}$ constitutes a microscopic continuum description of the experimental system, which greatly facilitates the analysis of the vortex pinning mechanism compared to the conventional lattice descriptions.

The corresponding Bogoliubov-de-Genne equation can be most naturally written in a basis expanded by the orbital angular momentum (OAM) partial waves and the Bessel functions (see more details in Part-II of supplementary materials). For $V_0 = 0$, the impurity is uncoupled to the SC. We calculate the in-gap spectrum that shown in Fig. 6(a). A series of in-gap CdGM states characterized by different OAMs take place. For $V_0 \neq 0$, the coupling term relevant to the in-gap physics can be readily derived from $H_{hyb}$ after the Bogoliubov transformation, i.e.

$$H_{\text{imp-CdGM}} = 2\pi \sum_\sigma \int dr r V_0(r)[u(r)d_\sigma^\dagger \gamma_{-1/2} + v^*(r)\gamma_{1/2}^\dagger d_\sigma] + h.c. \qquad (1)$$

where $u(r), v(r)$ are factors associated with the transformation. $\gamma_{\pm 1/2}$ denotes the Bogoliubov quasi-particle operator corresponding to the two lowest CdGM states with OAM

$m = \pm 1/2$, as marked by red in Fig. 6(a). Eq. (1) indicates that the major effect of the impurity is to couple with the lowest two CdGM states, as schematically depicted by Fig. 6(b). This impurity-CdGM coupling is the driving force for the shift of the CdGM states.

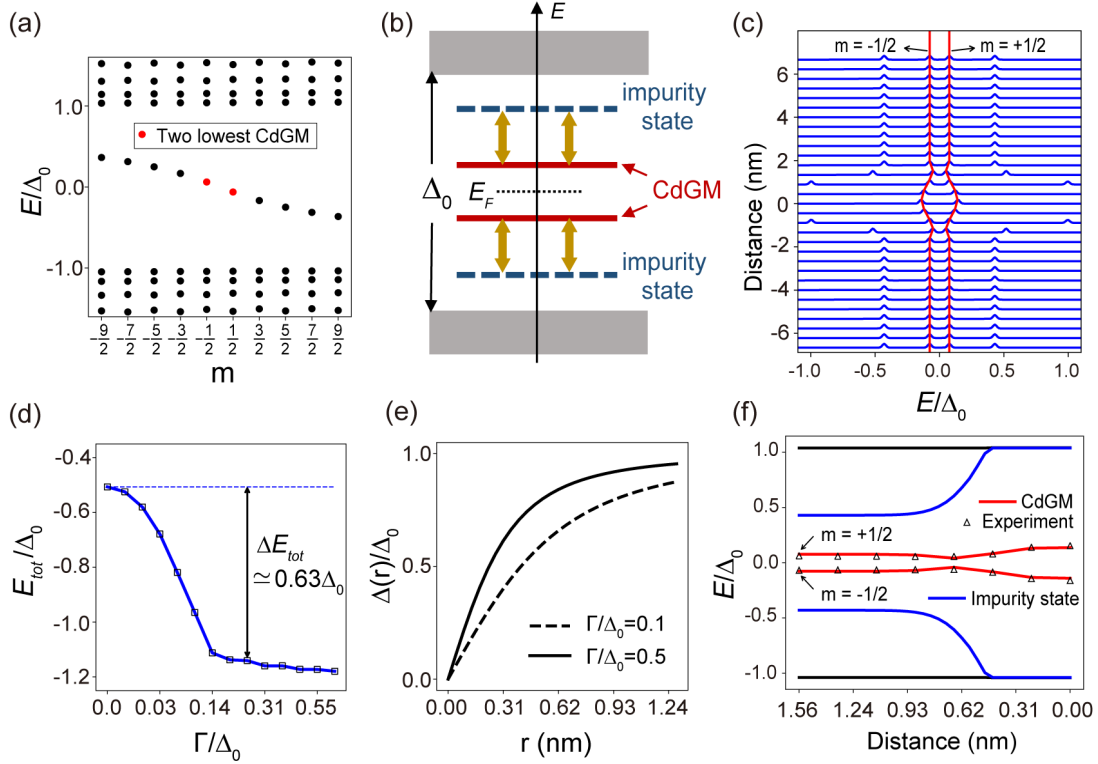

**FIG. 6.** Microscopic modeling of vortex pinning. (a) The calculated energy spectrum for the free vortex case. Two lowest CdGM states with OAM $m = \pm 1/2$ are marked by red. (b) The microscopic mechanism driving the shift of CdGM states. The impurity is strongly coupled to the lowest CdGM with OAM $m = \pm 1/2$, while leaves the other CdGM states nearly unchanged. (c) The evolution of the in-gap LDOS with varying $d$ firstly towards and then away from the vortex core. The red curve highlights the evolution of CdGM states. The two higher energy peaks are contributed by the impurity states. (d) The total energy of the lowest CdGM states and the impurity state as a function of $\Gamma$. (e) The local gap function, $\Delta(r)$, self-consistently determined for different $\Gamma$. (f) The zoom-in data in the region near the vortex core, which makes a closer comparison between the numerical results and the experiments. To better describe experiments where the dumbbell defect has a p-wave anisotropy, we added a small anisotropic scattering term of the strength $V_1$ (See Part II of supplementary materials). This term is not necessary for the general pinning mechanism and does not affect $\Delta E_{tot}$ obtained in panel (d), but it ensures the avoided level crossing between two lowest CdGM states with $m = \pm 1/2$. The parameters used in the calculations are: $\Delta_0$=5.5 meV, $\varepsilon_d = 2.4$ meV, $\mu = 110$ meV, $V_1 = 5.5$ meV, $r_0 = 0.53$ nm.

The hybridization $V_0(d)$ between the impurity and the local electrons decays with the distance to impurity ($d$). We calculated the evolution of the lowest CdGM states and the impurity states as $d$ changes. As illustrated in Fig. 6(c), when $d$ is much larger than $\xi$, the in-gap states remain unchanged. As approaching the vortex core ($d \lesssim 2$nm), due to the enhanced impurity coupling, the two lowest CdGM states are pushed towards slightly and then away from the Fermi energy $E_F$ (see part II-2 of supplementary materials for details). The zoom-in figure in Fig. 6(f) clearly shows a significant energy shift of the lowest CdGM states, which saturates at higher energies away from $E_F$ at $d=0$. These results are in quantitative agreement with the experimental data shown in Fig. 2(g). We also found that, accompanied by the shift of the CdGM states, the local pairing around the vortex core is simultaneously enhanced, as shown in Fig. 6(e). This enhancement is reflected by a reduced coherence length in the vortex core with an impurity, obtained by the self-consistent calculation of gap equation (see Part II-3 of supplementary materials).

The shift of the in-gap states inevitably modifies the energetics of the superconducting system. We then calculate and plot the total energy of the lowest CdGM states and the impurity state (below $E_F$), $E_{tot}$, as a function of the broadening function $\Gamma = \pi \rho_0 V_0^2$ with $\rho_0$ the DOS of the normal state. Since all the other CdGM states ($m \neq \pm 1/2$) and the above-gap continuum are barely affected by the impurity, the quantity $\Delta E_{tot} = E_{\text{tot}}(V_0) - E_{\text{tot}}(0)$ (for large $V_0$), is essentially the energy difference between a pinned vortex and a free vortex. Thus, $\Delta E_{tot}$ offers the magnitude of elementary pinning energy $U_{pin}$ in the perspective of a microscopic description. $\Delta E_{tot}$ is evaluated around $-0.63\Delta_0$ in Fig. 6(d), i.e., $-3.47$ meV for $\Delta_0 = 5.5$ meV [see Fig. 2(d)], which is of comparable value with the $U_{pin}$ estimated by experiments.

## IV. DISCUSSION AND CONCLUSION

So far, we have performed a comprehensive microscopic study on the elementary vortex pinning in FeSe-based superconductors. Our theoretical model well accounts for the experimental findings. We note the CdGM-impurity coupling revealed in Eq. (1) is a general result, which does not rely on the pairing details. As demonstrated in the supplementary material (Part II-5 of supplementary materials), for d-wave superconductors, the impurity is still able to shift the vortex-induced excitations, although there are no discrete bound states in the vortices. Besides, such an impurity-vortex coupling scenario does not rely on the impurity types, e.g., magnetic or non-magnetic. Although non-magnetic impurity does not affect the bulk property of isotropic s-wave superconductors, its effect is non-negligible when a vortex is in presence. This is because the impurity-pinned vortex locally breaks time-reversal symmetry (TRS), in analogy with magnetic impurities. This will inevitably lead to significant modifications to the in-gap physics, resulting in the pinning effect according to our calculations. In previous phenomenological pinning theory [1,2], pair breaking/non-breaking impurities are believed to affect the coefficient of different terms of G-L free-energy (so called $\delta T_C$ and $\delta l$ pinning). Here, we reveal that their microscopic pinning mechanism could be unified by the impurity-CdGM coupling. In addition, Fig. 6(d) shows that the impurity coupling strength $V_0$ is the key factor that determines the pinning energy and thus the pinning force. This observation may provide a guidance for the search of superconductors with large critical current $J_C$.

In summary, our tunneling spectroscopy measurement reveals the coupling between vortex and impurity state is the origin of vortex pinning. The elementary pinning energy/force are extracted from local tunneling spectra, which set up a direct connection between the microscopic electronic structure of vortex and the macroscopic transport quantity of $J_C$ for the first time. The pinning mechanism is well captured by our quantum impurity model in an s-wave superconductor with a vortex. The obtained results are not sensitive to the detailed features of the pairing and impurity, thus are applicable to broad classes of superconductors and pinning centers. Therefore, our study established a general microscopic mechanism of vortex pinning, the essential factor enabling practical superconductors to carry non-dissipative current.

## ACKNOWLEDGMENTS

We acknowledge the 1st National Conference of Magnetic Flux in Superconductors (Kunming, China 2023) organized by Prof. H. H. Wen and Prof. Z. X. Shi, which inspired this research. We also thank Prof. Q. H. Wang and Prof. Z. X. Zhao for helpful discussion. This work is supported by National Natural Science Foundation of China (Grants Nos.: 12225403, 92065202, 11888101, 92365302, 12322402, 12274206, 12304181, 12104094), Innovation Program for Quantum Science and Technology (Grant no.: 2021ZD0302800), The New Cornerstone Science Foundation (China), National Key Research and Development Program of China (Grant Nos. 2022YFA1403900, 2023YFA1406100), Key Research Program of Frontier Sciences of the Chinese Academy of Sciences (Grant No. XDB33010200), Science Challenge Project (Grant No. TZ2016004), Shanghai Municipal Science and Technology Major Project (Grant No. 2019SHZDZX01), Shanghai Pilot Program for Basic Research, Fudan University 21TQ1400100 (21TQ005), Xiaomi Foundation.

## APPENDIX: METHODS

**Sample preparation:** High quality $(Li_{0.8}Fe_{0.2})OHFeSe$ single crystalline films were grown on $LaAlO_3$ substrate by matrix-assisted hydrothermal epitaxy, as described in Refs. [27,28]. Single crystal $K_{0.8}Fe_{1.6}Se_2$ were used as precursor (matrix) which facilitated the hydrothermal epitaxial growth of $(Li_{1-x}Fe_x)OHFe_{1-y}Se$ films, resulting in single crystalline films with 100 - 400 nm thickness. Fig. 7 shows the XRD characterizations of the film. The observation of only $(00l)$ reflections [Fig. 7(a)] indicates its single preferred in plane orientation. Additional peaks marked with LAO are from the substrate. The x-ray rocking curve for the (006) Bragg reflection [Fig. 7(b)] has a FWHM of only 0.13°. The $\phi$-scan of (101) plane in Fig. 7(c) exhibits four successive peaks with an equal interval of 90°, consistent with the C4 symmetry of the film. This evidences an excellent out-of-plane orientation and epitaxial growth. The films were cleaved under ultra-high vacuum and transferred to STM head right after cleaving. 1ML FeSe/$SrTiO_3$ sample was grown by co-deposition of high purity Se (99.999%) and Fe (99.995%) on $SrTiO_3$ (001) substrate holding at 670 K, followed by annealing at ~800 K for 1 hours. The $SrTiO_3$ (001) (0.5% Nb doping) substrate were cleaned by direct heating at 1250K in ultra-high vacuum.

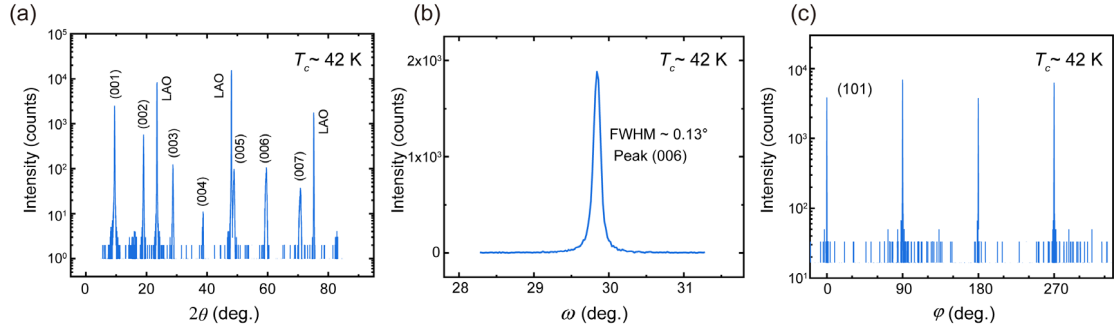

**FIG. 7** XRD characterizations of (Li,Fe)OHFeSe film on LaAlO$_3$. (a) The θ–2θ scan shows only (00*l*) peaks. (b) The X-ray rocking curve for the (006) peak with a full width at half maximum (FWHM) of 0.13°. (c) The φ-scan of the (101) plane.

**STM measurement:** The STM experiment was performed in a dilution refrigerator STM (Unisoku) at the base temperature of 20 mK ($T_{eff}$ = 160 mK) or at 4.2K when specified. Normal PtIr tips were used and cleaned by e-beam heating. Topographic images are taken with constant current mode with bias voltage ($V_b$) applied to the sample. The tunneling conductance $dI/dV$ is collected by standard lock-in method with a modulation frequency of 973 Hz. The typical modulation amplitude (ΔV) is 30 μeV at T = 20mK and 0.2 meV at T=4.2K.

**Transport measurement of $J_C$:** Transport measurements under magnetic field up to B = 14 T were carried out via standard four-probe method in a Quantum Design PPMS DynaCool system. The values of $J_C$ were obtained using the criteria of 1 μV on I–V curves and the bridge parameters were characterized by a Bruker DektakXT stylus profilometer.

**Solving the Bogoliubov-de-Genne (BdG) equation of superconductors with a pinned vortex:** Three steps of transformation are performed to cast the BdG Hamiltonian in a proper matrix form. We first make a gauge transformation that removes the phase winding of the pairing potential. Then, we expand the wave function in terms of the partial waves of different OAMs, denoted by *m*. Finally, the radial dependence of the wave function is expanded in the complete basis formed by Bessel functions. For each *m*, the final BdG Hamiltonian is written into a (4*N* +4) dimensional matrix, where the 4*N* comes from tensor product of the Nambu, spin and the Bessel function space. *N* is the cutoff in terms of the number of Bessel functions used in the expansion, and *N*=100 is used in the calculations, which is large enough to ensure numerical convergence. The additional 4 dimension of the matrix comes from the impurity Hilbert space, which consists of the empty, doubly-occupied state, and two singly-occupied states. Exact diagonalization then generates a complete and accurate spectrum of the experimental system, including the CdGM, the impurity state, and the above-gap continuum. More details are included in Part II of the supplemental materials.

# Supplementary Materials for
## "Revealing the Microscopic Mechanism of Elementary Vortex Pinning in Superconductors"

C. Chen†, Y. Liu†, Y. Chen†, Y. N. Hu, T. Z. Zhang, D. Li, X. Wang, C. X. Wang, Z.Y.W. Lu, Y. H. Zhang, Q. L. Zhang, X. L. Dong, R. Wang*, D. L. Feng*, T. Zhang*

†These authors contributed equally.
*Corresponding authors: rwang89@nju.edu.cn, dlfeng@ustc.edu.cn, tzhang18@fudan.edu.cn

**Part I: Additional data and the determination of elementary pinning energy and pinning force.**

**I -1. Additional experimental data:**

Besides the dumbbell-like defect at the Fe-site, there are Se vacancies in the FeSe surface of (Li,Fe)OHFeSe, as shown in Fig. S1(a). These defects do not show observable pinning effect on the vortices. Fig. S1(b) displays a vortex core appeared in the region of Fig. S1(a). It overlapped with two Se vacancies but no obvious influence on the vortex state is observed.

Fig. S2 shows additional vortex maps of (Li,Fe)OHFeSe and dI/dV spectra taken at 9 pinned vortices and 3 free vortices. All the pinned vortex cores display a mini-gap at $E_F$. The half-width of the gap (the lowest energy CdGM state) is in the range of 0.8~1.6 meV. Meanwhile the free vortex cores all display zero-bias peaks without mini-gap.

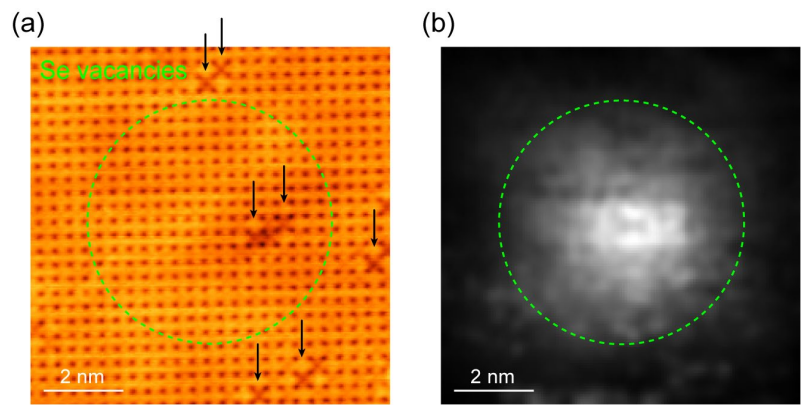

**FIG. S1.** (a) Topographic image of the FeSe surface of (Li,Fe)OHFeSe, showing Se vacancies (indicated by black arrows). ($V_b$ = 30mV, $I$ = 60pA). (b) Zero-bias dI/dV map taken at the region of panel (a) at B=8.5T, showing a vortex core ($V_b$ = 30mV, $I$ = 60pA).

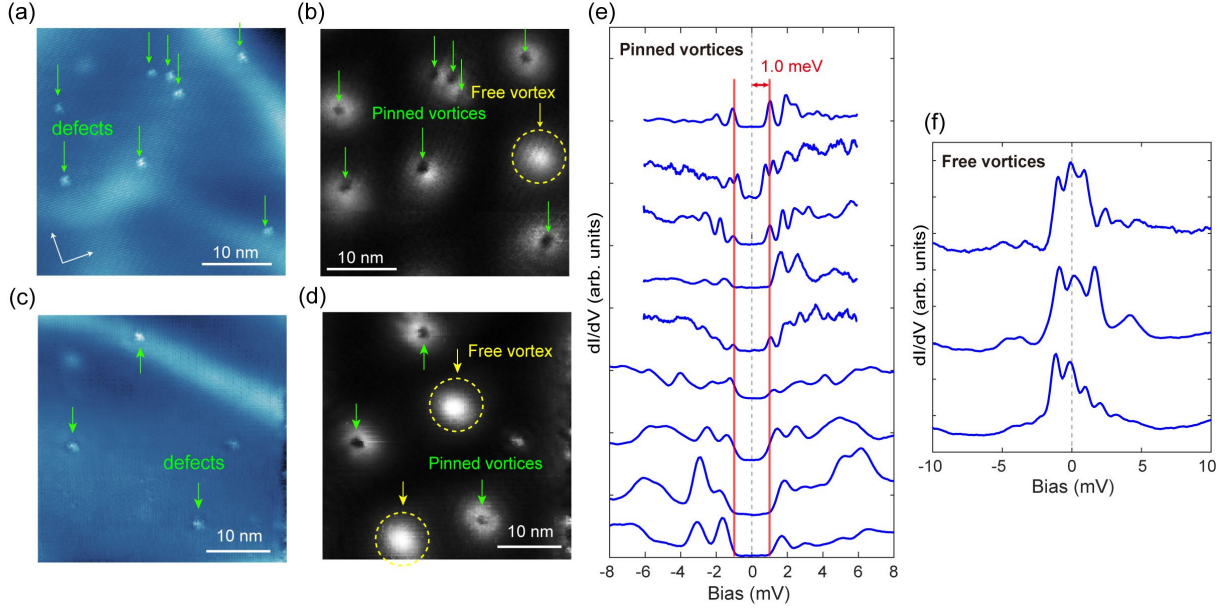

**FIG. S2.** (a,c) Two surface region of (Li,Fe)OHFeSe, with green arrows indicating dumbbell defects. (b,d) Zero-bias dI/dV map taken at the regions of panel (a,c) under B = 10T, respectively ($V_b$ = -30mV, $I$ = 40pA). Green arrows indicate pinned vortices and yellow arrows indicate free vortices. (e,f) dI/dV spectra taken at the core center of different pinned & free vortices. The size of mini-gap is in the range of 0.8~1.6 meV. The red line indicates the energy of 1.0 meV.

## I -2. Extracting the elementary pinning energy through dI/dV linecut spectra:

In STM measurement, the tunneling conductance (dI/dV) at a bias voltage of V is proportional to the local density-of-state (LDOS) of the sample, assuming the tip DOS and tunneling matrix element are constants near $E_F$:

$$\frac{dI}{dV}(V, \boldsymbol{r}) \propto \rho_s(E_F + eV, \boldsymbol{r})$$

In a superconducting state, the energy gap opening in single-particle DOS lowers the free-energy of the system, which contributes to the main part of condensation energy. Any in-gap bound state (like the CdGM state) will reduce the condensation energy and be destructive to superconductivity. Via integrating the LDOS over an energy range covers superconducting gap and an area covers the whole pinned/free vortex core, we can obtain the condensation energy difference between a pinned and a free vortex core, which is defined as the elementary pinning energy $U_{pin}$. The absolute value of LDOS can be obtained by calibrating the dI/dV spectra with the normal state DOS near $E_F$ of the host superconductor. Then $U_{pin}$ is expressed as:

$$U_{pin} = \int \int_{E_{LB}}^{E_F} N(0) \left[ \left(\frac{dI}{dV}\right)_{pinned} - \left(\frac{dI}{dV}\right)_{free} \right] E dE ds$$

Here the d$I$/d$V$ spectra are normalized by their value at the energy of $E_{LB}$ = -17 meV, which is outside of superconducting gap and the d$I$/d$V$ curve is nearly flat at $E_{LB}$, as shown in Fig. S3. $N(0)$ is the normal state DOS near $E_F$ (per area). For (Li,Fe)OHFeSe and 1 ML FeSe/SrTiO$_3$ superconductors, their Fermi surfaces (contributed by a single FeSe layer) are composed of *two* near circular electron pockets at the M point (refs. *32-34*), as sketched in Fig. S4(a). $N(0)$ can be extracted from the band dispersion of these pockets measured by *in-situ* quasi-particle interference (QPI) shown Fig. S4. Here we used a parabolic curve $E(q) = aq^2 - E_b$ ($q = 2k$) to fit the dispersion in Fig. S4(d,e). which yield $E_b = 50$meV, $k_F = 0.17$Å$^{-1}$ for (Li,Fe)OHFeSe, and $E_b = 60$meV, $k_F = 0.20$ Å$^{-1}$ for 1ML FeSe/SrTiO$_3$. Then $N(0) = 2S \cdot 2\pi k dk/4\pi^2 dE = 2S k_F^2/4\pi E_b$ ($S=2$ is the spin degeneracy).

For free vortex and a pinned vortex with a defect at its center, their CdGM states all have near isotopic distribution. Then the spatial integration of LDOS can be obtained through a linecut dI/dV spectra taken across the vortex core (as shown in Figs. S3(a,b,d,e)), via: $\int (\frac{dI}{dV})_{(r)} 2\pi r dr$. All the integrations were performed numerically with a spatial resolution <0.5 nm and an energy resolution <0.1 meV. The calculated value of $U_{pin}$ is -1.8 meV for (Li,Fe)OHFeSe and -2.3 meV for 1 ML FeSe/SrTiO$_3$.

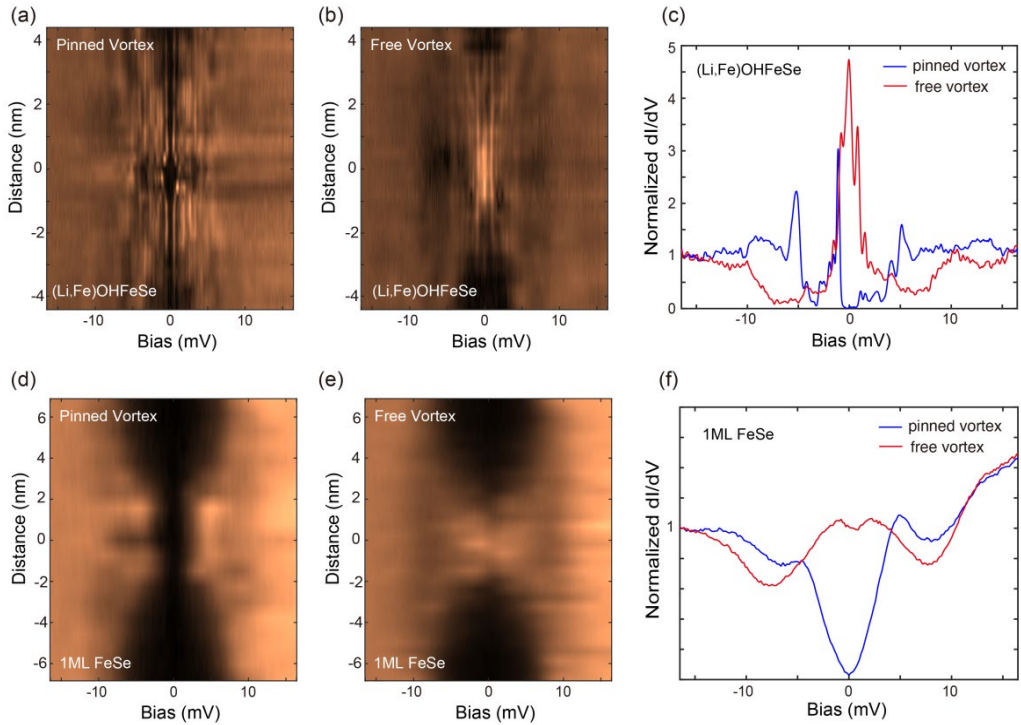

**FIG. S3.** (a,b) Color plots of linecut dI/dV spectra taken across the center of pinned vortex and free vortex in (Li,FeOH)FeSe. ($V_b = 17$mV, $I = 60$pA, $T_{eff} = 160$mK) (c) Normalized dI/dV spectra taken at the center of pinned and free vortex cores of (Li,Fe)OHFeSe. (d, e) Color plots of linecut dI/dV spectra taken across the center of pinned vortex and free vortex in 1ML FeSe/SrTiO$_3$ ($V_b = 30$mV, $I = 80$pA, T = 4.2K). (f) Normalized dI/dV spectra taken at the center of pinned and free vortex cores of 1ML FeSe.

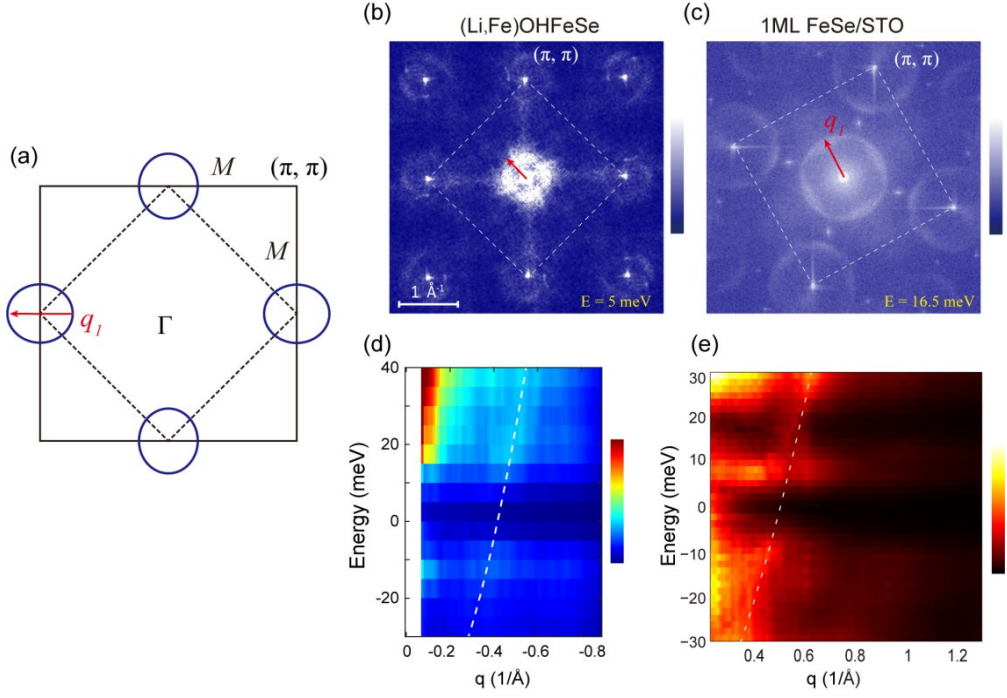

**FIG. S4.** (a) Schematic of the (unfolded) Brillouin zone (BZ) and Fermi surface of (Li,Fe)OHFeSe and single-layer FeSe/SrTiO$_3$(001). (b,c) The representative QPI pattern (FFT image) of (Li,Fe)OHFeSe and 1ML FeSe/SrTiO$_3$(001), measured at E= 5 meV and E = 16.5 meV (respectively). (d,e) The Q-space dispersion of the electron pocket of (Li,Fe)OHFeSe and 1ML FeSe/SrTiO$_3$(001), respectively. Dashed curves are parabolic fittings using $E = \frac{E_b}{q_F^2}q^2 - E_b$, which yield $E_b$ = 50 meV, $q_F$ = 0.35 Å$^{-1}$ for (Li,Fe)OHFeSe, and $E_b$ = 60 meV, $q_F$ = 0.40 Å$^{-1}$ for 1ML FeSe/SrTiO$_3$.

## I- 3. Calculation of pinning force through the spatial variation of pinning energy

As illustrate in Fig. 4(a), the pinning force $f_p$ can be calculated through the spatial gradient of $U_{pin}$ when the pinned vortex is forced to leave the pinning site (defect site). We calculated the $U_{pin}$ difference between a pinned vortex measure at 7T and 6T, as shown in Fig. S5. The $U_{pin}$ can be calculated by the similar method shown above, except that here the LDOS distribution is no longer isotropic (since the vortex is shifted away from the defect). Therefore, we used full dI/dV maps taken over the vortex at various energies to do the spatial integration of LDOS, as shown in Fig. S5 below. The absolute value of LDOS is also calibrated in the similar way shown above. To increase the accuracy of the results, considering the superconducting DOS has particle-hole symmetry, the dI/dV was symmetrized with respect to zero-bias before integration.

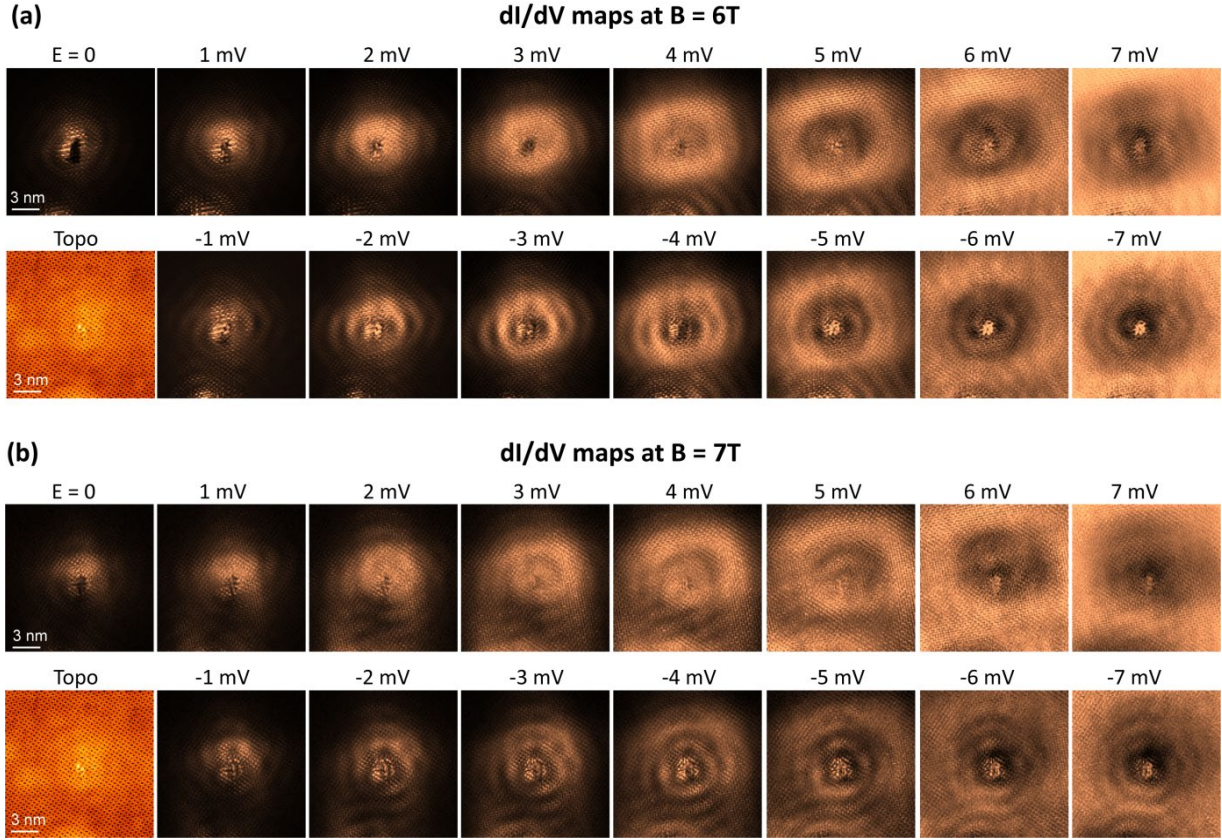

**FIG. S5.** (a) dI/dV maps of the pinned vortex taken at different energy, under B = 6T (setpoint: $V_b$ = 10mV, $I$ = 60pA, $T_{eff}$ = 160mK). (b) dI/dV maps of the pinned vortex taken at different energy, under B = 7T (setpoint: $V_b$ = 10mV, $I$ = 60pA, $T_{eff}$ = 160mK).

**Part II: Microscopic modeling of the vortex CdGM state hybridized with the impurity state.**

**II -1. Model and method**

The SC system under our investigation consists of a local impurity and a vortex, which are bound to each other. We start from normal state electrons with attractive interaction induced by electron phonon interaction. The BCS-reduced Hamiltonian in two-dimensions (2D) reads as $H_{BCS} = \sum_{k\sigma} \varepsilon_{k\sigma} c^\dagger_{k\sigma} c_{k\sigma} - (U/2) \sum_{k,k',\sigma,\sigma'} c^\dagger_{k',\sigma} c^\dagger_{-k',-\sigma} c_{-k,-\sigma} c_{k,\sigma}$, where $\varepsilon_{k\sigma} = k^2/2m - \mu$ with $m$ being the electron mass and $\mu$ the chemical potential of the normal state. At mean-field level and in the region away from the impurity and vortex, a uniform order parameter $\Delta_0$ can be formed, leading to the conventional self-consistent equation for s-wave superconductors (SCs), i.e., $1 = (U/2) \sum_k 1/\sqrt{\varepsilon_{k\sigma}^2 + \Delta_0^2}$. Whereas, near the impurity and vortex, the spatial variation and the phase winding of the order parameter should be considered. The impurity its coupling to the SC also needs to be considered. Thus, the system is described by the quantum impurity model, $H = H_{\text{SC}} + H_{\text{imp}} + H_{\text{hyb}}$, where $H_{\text{SC}}$ is the mean-field BCS Hamiltonian with spatially varying order parameter, i.e.,

$$H_{\text{SC}} = \sum_\sigma \int d\mathbf{r} \left[ -\frac{\hbar^2}{2m}\nabla^2 - \mu \right] c^\dagger_{\mathbf{r}\sigma} c_{\mathbf{r}\sigma} + \int d\mathbf{r} [\Delta(\mathbf{r}) c^\dagger_{\mathbf{r}\uparrow} c^\dagger_{\mathbf{r}\downarrow} + h.c.], \quad (1)$$

To incorporate the effect of a magnetic flux in the SC at $\mathbf{r} = 0$, we let $\Delta(\mathbf{r}) = \Delta(r) e^{i\nu\theta}$, where the radial dependence $\Delta(r) = \Delta_0 r/\sqrt{r^2 + \xi^2}$ is assumed (*37*), with $\xi$ being the SC coherent length at the vortex, and $\nu$ an integer winding number. Clearly, $\Delta(r)$ vanishes at the origin $\mathbf{r} = 0$ and reaches $\Delta_0$ for $r \gg \xi$. Meanwhile, the phase of the pairing potential goes from 0 to $2\pi\nu$ around $\mathbf{r} = 0$ with $\theta$ going from 0 to $2\pi$. In the following, we consider the case where the vortex carries one flux quantum, i.e., $\nu = 1$ is assumed. Here, the uniform order parameter $\Delta_0$ and $\xi$ are treated as parameters that are self-consistently determined in mean-field level.

Here, we consider the normal s-wave superconductor. It will be clear in the following that the shifting behavior of the CdGM modes remains qualitatively the same for other pairing symmetries, such as the p+ip pairing. Moreover, the shifting behavior of the CdGM modes does not rely much on whether the impurity is magnetic or non-magnetic. Thus, we consider a non-magnetic quantum impurity model described by

$$H_{\text{imp}} = \sum_\sigma \epsilon_d d^\dagger_\sigma d_\sigma, \quad (2)$$

where $d^\dagger_\sigma (d_\sigma)$ is the creation (annihilation) operator of the local impurity state with spin $\sigma$. Furthermore, the hybridization between the impurity and the electrons of the SC state is given by

$$H_{\text{hyb}} = \sum_\sigma \int d\mathbf{r} [V_0(\mathbf{r}) d^\dagger_\sigma c_{\mathbf{r}\sigma} + h.c.], \quad (3)$$

where $V_0(\mathbf{r})$ is the scattering strength between the electrons in SC and the impurity, which usually takes the following form:

$$V_0(\mathbf{r}) = V_0 \frac{1}{\sqrt{\pi} r_0} e^{-(r/r_0)^2}, \quad (4)$$

where $r_0$ is the decaying length.

The Eqs. (1)-(3) above constitute a non-interacting Anderson-type quantum impurity model in a s-wave superconductor along with a magnetic vortex. The vortex center is located at the impurity position $\mathbf{r} = 0$ due to the pinning effect. It is also important to note that, the impurity in (LiFe)OHFeSe has a dumbbell shape that is of p-wave anisotropy (*25*). This brings about anisotropic scattering of the electrons in the SC. Thus, we further consider an anisotropic potential seen by the electrons, which is described by:

$$H_{\text{ani}} = \sum_\sigma \int d\mathbf{r} V_1(\mathbf{r}) c^\dagger_{\mathbf{r}\sigma} c_{\mathbf{r}\sigma}, \quad (5)$$

where $V_1(\mathbf{r}) = 4V_1 \cos^2(\theta/2) e^{-(r/r_0)^2}/(\sqrt{\pi} r_0)$ is the strength of the p-wave scattering potential seen by the electrons around the impurity, and $\theta$ is the polar angle of $\mathbf{r}$. $V_1$ generates anisotropy, yet its value is dominated by the isotropic component $V_0$ in the experimental case. It is straightforward to include higher order scattering component in Eq.(5), however the shifting behavior of the CdGM mode remains qualitatively the same.

For the vortex-free case with $\nu=0$, the SC Hamiltonian $H_{SC}$ can be conveniently diagonalized by Bogoliubov transformation,

$$\gamma_n = \int d\mathbf{r} \sum_\sigma [u_{n\sigma}^*(\mathbf{r})c_{\mathbf{r}\sigma} + v_{n\sigma}^*(\mathbf{r})c_{\mathbf{r}\sigma}^\dagger],$$

$$\gamma_n^\dagger = \int d\mathbf{r} \sum_\sigma [u_{n\sigma}(\mathbf{r})c_{\mathbf{r}\sigma}^\dagger + v_{n\sigma}(\mathbf{r})c_{\mathbf{r}\sigma}],$$

(6)

where $\gamma_n^\dagger(\gamma_n)$ is the creation (annihilation) operator of the Bogoliubov quasi-particles. The Bogoliubov-de Gennes (BdG) equation leads to the eigenvalues $E_n$ in the diagonal basis $\Phi_n(\mathbf{r}) = [u_{n\uparrow}(\mathbf{r}), u_{n\downarrow}(\mathbf{r}), v_{n\downarrow}(\mathbf{r}), -v_{n\uparrow}(\mathbf{r})]^T$, i.e.,

$$h_0(\mathbf{r})\Phi_n(\mathbf{r}) = E_n \Phi_n(\mathbf{r}), \tag{7}$$

where $h_0(\mathbf{r})$ is the single-particle Hamiltonian of the SC, i.e.,

$$h_0(\mathbf{r}) = \begin{bmatrix} -\dfrac{\hbar^2}{2m}\nabla^2 - \mu & 0 & \Delta(\mathbf{r}) & 0 \\ 0 & -\dfrac{\hbar^2}{2m}\nabla^2 - \mu & 0 & \Delta(\mathbf{r}) \\ \Delta^*(\mathbf{r}) & 0 & \dfrac{\hbar^2}{2m}\nabla^2 + \mu & 0 \\ 0 & \Delta^*(\mathbf{r}) & 0 & \dfrac{\hbar^2}{2m}\nabla^2 + \mu \end{bmatrix}, \tag{8}$$

In the presence of a vortex, $\nu=1$, we make a gauge transformation $\Phi_n(\mathbf{r}) \to \tilde{\Phi}_n(\mathbf{r}) = e^{-i\nu\theta\tau_z/2}\Phi_n(\mathbf{r})$ where $\tau_z$ is the Pauli matrix in the Nambu space. Under the gauge transformation, the phase of the pairing potential is effectively removed, i.e., $\Delta(\mathbf{r}) \to \tilde{\Delta}(\mathbf{r}) = e^{-i\nu\theta}\Delta(\mathbf{r}) = \Delta(r)$. Thus, the transformation in terms of the basis reads as, $\Phi_n(\mathbf{r}) = e^{i\nu\theta\tau_z/2}\tilde{\Phi}_n(\mathbf{r}) = e^{i\nu\theta\tau_z/2}[u_{n\uparrow}(\mathbf{r}), u_{n\downarrow}(\mathbf{r}), v_{n\downarrow}(\mathbf{r}), -v_{n\uparrow}(\mathbf{r})]^T$. Then, we expand the wave function to the orbital angular momentum (OAM) space via $\Phi_n(\mathbf{r}) = \Sigma_m e^{-im\theta}\Phi_{nm}(r)$, where

$$\Phi_{nm}(r,\theta) = \int_0^{2\pi} d\theta\, \Phi_n(\mathbf{r}) = [u_{n,m+\nu/2\uparrow}(r), u_{n,m+\nu/2\downarrow}(r), v_{n,m-\nu/2\downarrow}(r), -v_{n,m-\nu/2\uparrow}(r)]^T.$$

(9)

Furthermore, to accurately solve the BdG equation, we expand the radial functions $u_{nm\sigma}(r)$, $v_{nm\sigma}(r)$ in the complete basis of Bessel functions, i.e.,

$$u_{nm\sigma}(r) = \sum_j u_{nmj\sigma}\phi_{mj}(r),$$

$$v_{nm\sigma}(r) = \sum_j v_{nmj\sigma}\phi_{mj}(r),$$

(10)

where

$$\phi_{mj}(r) = \frac{\sqrt{2}}{R J_{m+1}(\beta_{mj})} J_m\left(\beta_{mj} \frac{r}{R}\right). \tag{11}$$

$J_m(r)$ is the m-th order Bessel function defined in a disc of radius $R$, and $\beta_{mj}$ is the j-th root of $J_m(r)$. The basis clearly satisfies the orthogonality, $\int_0^R dr r \phi^*_{mj'}(r) \phi_{mj}(r) = \delta_{jj'}$. Under the above basis, the BdG equation is finally cast into:

$$\begin{bmatrix} T_{m+\nu/2} & 0 & \Delta_{m+\nu/2,m-\nu/2} & 0 \\ 0 & T_{m+\nu/2} & 0 & \Delta_{m+\nu/2,m-\nu/2} \\ \Delta^t_{m+\nu/2,m-\nu/2} & 0 & -T_{m-\nu/2} & 0 \\ 0 & \Delta^t_{m+\nu/2,m-\nu/2} & 0 & -T_{m-\nu/2} \end{bmatrix} \begin{bmatrix} u_{nj,m+\nu/2\uparrow} \\ u_{nj,m+\nu/2\downarrow} \\ v_{nj,m-\nu/2\downarrow} \\ -v_{nj,m-\nu/2\uparrow} \end{bmatrix} = E_n^m \begin{bmatrix} u_{nj,m+\nu/2\uparrow} \\ u_{nj,m+\nu/2\downarrow} \\ v_{nj,m-\nu/2\downarrow} \\ -v_{nj,m-\nu/2\uparrow} \end{bmatrix},$$
(12)

where $T_m$ and $\Delta_{m+\nu/2,m-\nu/2}$ are matrices in the Bessel function basis with the entries:

$$(T_m)_{ij} = -\left[\frac{1}{2m}\left(\frac{\beta_{mj}}{R}\right)^2 + \mu\right]\delta_{ij},$$
$$(\Delta_{mn})_{ij} = \int_0^R dr r \Delta(r) \phi_{mi}(r) \phi_{nj}(r). \tag{13}$$

The above transformation to the Nambu-OAM-Bessel basis can be written compactly in the second quantized form, namely, the original electron operators are transformed to the Bogoliubov quasiparticle operators via,

$$\gamma^\dagger_{nm} = \frac{1}{2\pi} \sum_\sigma \int d\mathbf{r} e^{im\theta} [u_{n,m+\nu/2,\sigma}(r) e^{i\nu\theta/2} c^\dagger_{\mathbf{r}\sigma} + v_{n,m-\nu/2,\sigma}(r) e^{-i\nu\theta/2} c_{\mathbf{r}\sigma}], \tag{14}$$

and

$$c_{\mathbf{r}\sigma} = \sum_{nm} [e^{im\theta} u_{n,m+\nu/2,\sigma} e^{i\nu\theta/2} \gamma_{nm} + e^{-im\theta} v_{n,m-\nu/2,\sigma} e^{i\nu\theta/2} \gamma^\dagger_{nm}]. \tag{15}$$

By firstly turning off the coupling to the impurity, we can readily obtain the OAM-resolved energy spectrum of the SC with a vortex by diagonalization of the Hamiltonian in Eq. (12). As shown in Fig. S6, the in-gap states in Fig. S6 clearly demonstrate the existence of CdGM modes carrying different OAM quantum numbers, which are half integers $m = l + \nu/2$ with integer $l \in Z$. The CdGM modes with different OAM $m$ have different energies, which are more and more away from zero energy with increasing $|m|$.

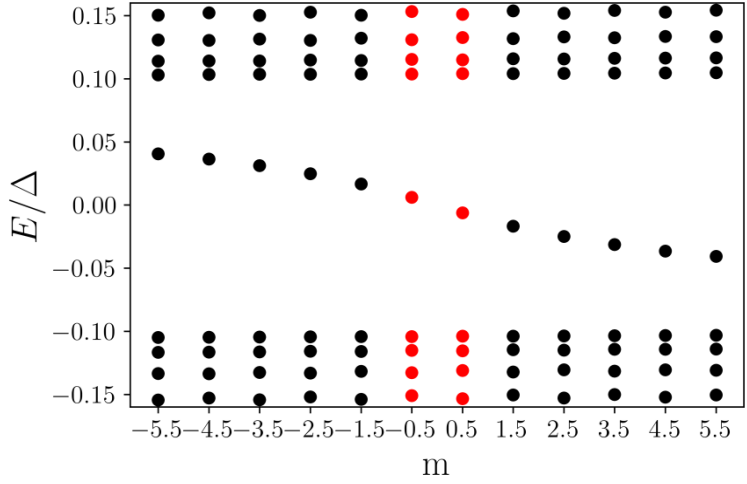

**FIG. S6.** The OAM-resolved energy spectrum with *m* from -11/2 to 11/2. The red dots highlight the spectra with the OAM $m= \pm 1/2$.

### II -2. The effective coupling between the CdGM and the quantum impurity.

Using Eq. (15) above, the hybridization term in Eq. (3) can be mapped to

$$H_{\text{hyb}} = 2\pi \sum_{n\sigma} \int dr r V(r) d_\sigma^\dagger [u_{n0\sigma}(r)\gamma_{n,-1/2} + v_{n0\sigma}^*(r)\gamma_{n,1/2}^\dagger] + h.c. \quad (16)$$

Clearly Eq. (16) describes the coupling between the quantum impurity and the Bogoliubov quasi-particles with the OAM $m = \pm 1/2$, where we have focused on the case $\nu = 1$. The Bogoliubov quasi-particles that are coupled to the quantum impurity are marked by red in the energy spectrum as shown in Fig. S6. It is clear from Eq.(16) that, except for the above-gap bulk states, the in-gap states that are effectively coupled to the impurity only involves the lowest energy CdGM modes ($n = 0$) with the OAM $m = \pm 1/2$. From Eq. (16), their effective coupling can be read off as:

$$H_{\text{imp-CdGM}} = 2\pi\alpha \sum_\sigma \int dr r V_0(r) [u_{00\sigma}(r) d_\sigma^\dagger \gamma_{0,-1/2} + v_{00\sigma}^*(r)\gamma_{0,1/2}^\dagger d_\sigma] + h.c., \quad (17)$$

where $V_0(r) = V_0 e^{-(r/r_0)^2}/(\sqrt{\pi} r_0)$. The mixing term above will inevitably shift the energy level of both the lowest CdGM mode and the quantum impurity state. In order to trace the effect of the impurity-CdGM coupling, we intentionally introduce a controlling factor $\alpha$ with $\alpha \in [0,1]$ in front of Eq. (17). For $\alpha = 0$, one artificially turns off the impurity-CdGM coupling, while $\alpha = 1$ reproduces the realistic case with the full coupling.

In terms of the impurity scattering term, we firstly consider the isotropic and short-range case by taking $V_1 = 0$ and $r_0 \to 0$. In this case, the Gaussian function in $V_0(r)$ approaches the delta-function $\delta(\mathbf{r})$. We numerically solve the Eq. (2), Eq.(12) and Eq. (16) in the Nambu-OAM-Bessel basis introduced above. In our calculation, we introduce a cutoff $N = 100$ in terms of the Bessel basis function, which is sufficient as we have numerically confirmed that the results are saturated and remain unchanged for even larger $N$.

Fig. S7(a)(b) shows the evolution of the energy levels of the in-gap states with increasing the broadening function $\Gamma = \pi \rho_0 V_0^2$, where $\rho_0$ is the density of states (DOS) of the normal state. As shown in Fig. S7, the behavior of the in-gap states with increasing $\Gamma$ strongly depends on the relative energies of the impurity state and the CdGM mode at zero coupling $V_0 = 0$. From Fig. S7(a) where $\epsilon_d < |E_{\text{CdGM},0}|$, we find that when the effective coupling between the two states is turned off with $\alpha = 0$, the impurity level and the lowest CdGM mode (with OAM $m = \pm 1/2$) crosses with each other with increasing $\Gamma$ (the dashed curves in Fig. S7(a)). Hence, under the full coupling with $\alpha = 1$, the two states strongly hybridize with each other, as shown by Fig. S7(a). This is further verified by the calculated ratio of the wave function amplitude of the two states, as shown by Fig. S7(c). However, in Fig. S7(b) where $\epsilon_d > |E_{\text{CdGM},0}|$, the two states do not cross with each other at $\alpha = 0$, as shown by the dashed curves. Thus, we do not observe a strong mixing of the two states. Instead, the coupling between the two states generates the shifting of the CdGM mode. Specifically, the impurity state at the electron (hole) side is pushed to higher (lower) energies while the $m = 1/2$ ($m = -1/2$) CdGM mode is pushed towards lower (higher) energies. Interestingly, the $m = 1/2$ and the $m = -1/2$ CdGM modes cross with each other at zero energy at a critical $\Gamma_c$, as shown in Fig. S7(b).

The evolution of the impurity state (on the electron side) and the $m = 1/2$ CdGM mode is shown in Fig. S7(d), with a fixed $\Gamma$ and an increasing $\alpha$. It is clearly shown that the impurity-CdGM hybridization has a "repulsion effect", which pushes the two states away from each other in energies. As will be clear in the following, this "repulsion effect" can well account for our experiments at the quantitative level.

It should now be clear that, since it is the coupling with the impurity that drives the shift of the CdGM state, and the coupling remains qualitatively unchanged for other pairing symmetries, e.g., the p+ip SC, the shifting behavior of the CdGM mode should be insensitive to the SC pairing symmetries. This further supports our starting point in Eq.(1).

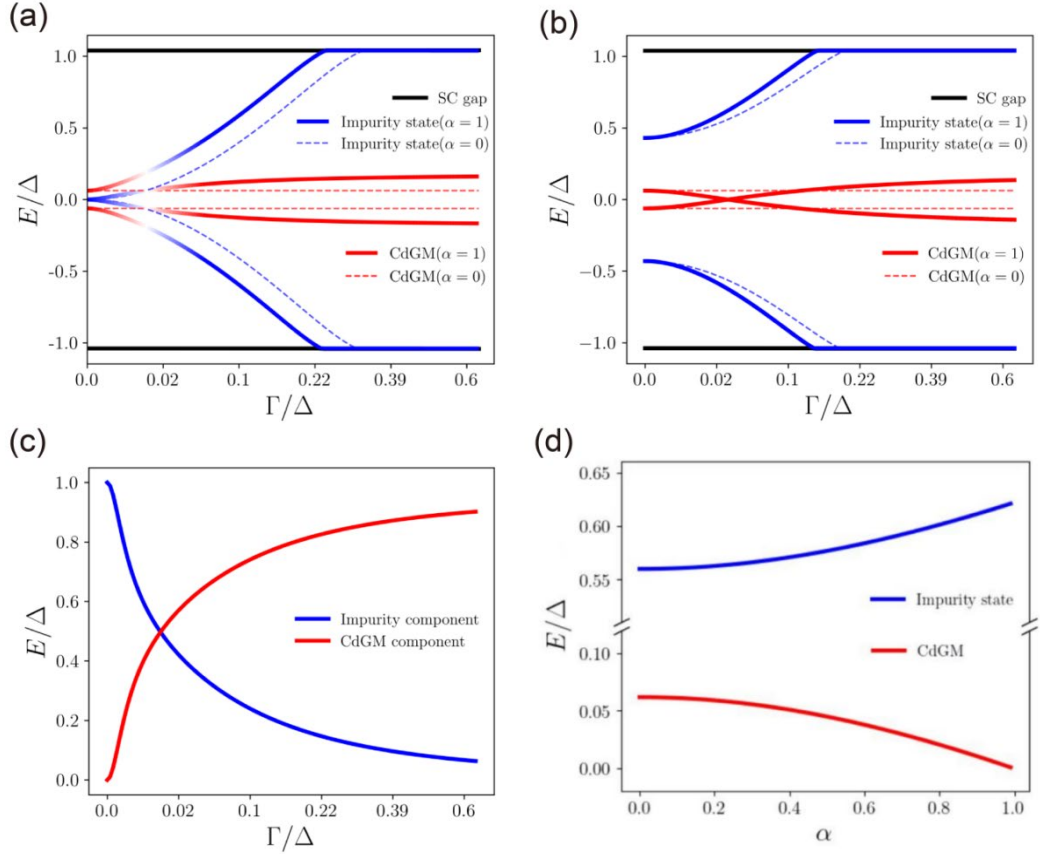

**FIG. S7.** The in-gap states calculated for the isotropic scattering $V_1 = 0$. (a,b) The in-gap states as a function of the broadening function $\Gamma$. The solid and dashed curve show the results for $\alpha = 1$ and 0, respectively. (c) The calculated ratio of the wave function amplitude between the impurity and the CdGM state as a function $\Gamma$. (d) The evolution of the impurity state (on the electron side) and the $m = 1/2$ CdGM state with varying α from 0 to 1. $\Gamma$ is set to the critical value $\Gamma_c$ where the $m = 1/2$ and $m = -1/2$ CdGM modes cross with each other in (b). $\epsilon_d = 0$ is used in (a,c) and $\epsilon_d = 0.043$ in (b,d).

## II -3. Self-consistent determination of the local pairing function around the vortex core.

In real space, the shift of the CdGM state is manifested by the localization around the vortex center. We calculate the spatial distribution of the two lowest CdGM states for different $\Gamma$. As shown by the Fig. S8(a),(b), the CdGM states become much more localized for the pinned vortex case. Thus, more states are squeezed towards the vortex center $r = 0$. Consequently, the local gap function $\Delta(r)$ around the vortex center is expected to be effectively enhanced.

The gap function $\Delta(r)$ around a free vortex should, in principle, be self-consistently determined. According to previous studies, e.g., Ref. 37, a satisfactory description of the gap function can take the form as $\Delta(r) = \Delta_0 r/\sqrt{r^2 + \xi^2}$. It should be noted that the shape of the gap profile is known to affect only slightly the quasi-particle energies; it does not change the key features of the vortex core states which are mainly controlled by the vortex topology. Therefore, the adopted function form of $\Delta(r)$ is expected to closely reproduce the true core spectrum and the corresponding eigenstates, for both the free vortex and the pinned vortex cases.

When the vortex is pinned by the defect, the gap functions $\Delta(r)$ will be quantitatively modified (without affecting the vortex topology). Since $\Delta_0$ is the gap away from the vortex, its value remains unchanged. Whereas, the local coherent length $\xi$ in $\Delta(r)$ can be modified. In order to investigate how the local pairing is changed by the defect, we treat $\xi$ as a variational parameter in the SC order parameter, and self-consistently determine its value at the mean-field level. Specifically, we calculate the total energy of the whole system,

$$E_{\text{sys}} = E_{\text{CdGM}} + E_{\text{imp}} + E_{\text{SC}} + \frac{\Delta_0^2}{U},$$

where $U$ is the attractive interaction between electrons induced by the electron-phonon coupling, $E_{SC}$ is the mean-field energy of all the Bogoliubov quasi-particles in the bulk continuum below the Fermi level, and $E_{CdGM}$ and $E_{imp}$ are the energy of the in-gap CdGM states and the impurity states, respectively, which are dependent on $\Delta(r)$ and thus on $\xi$. The SC gap away from the vortex $\Delta_0$ is determined by the conventional gap equation $1 = \frac{1}{2} U \sum_k \frac{1}{\xi_k}$, where $\xi_k = \sqrt{(\frac{k^2}{2m} - \mu)^2 + \Delta_0^2}$. Through minimizing $E_{sys}$, we are able to determine $\xi$ for different impurity coupling $V_0$ (and fixed $U$). The results are shown by Fig. 6(e) of the main text, where we observe that $\xi$ decreases with increasing $V_0$ or $\Gamma$. Hence, compared to the free vortex case, the local pairing potential is effective enhanced when an impurity is in presence.

We also plot the calculated LDOS at the vortex center for different parameters in Fig. S8(c),(d), with taking into account the contribution from all the CdGM states, the impurity state and the above-gap continuum. For a larger broadening factor, we observe from Fig. S8(d) that the SC gap (the black curve) is effectively obscured by the vortex due to the in-gap CdGM states (the blue curve), which is however enhanced when one further considers the pinning effect of the impurity (the red curve). For a smaller broadening factor, the in-gap peaks are more clearly shown, as displayed by Fig. S8(c). We note Fig. S8(c) and (d) are in qualitative agreement from the experimental data in Fig. 2(f) and Fig. 3(e) of the main text.

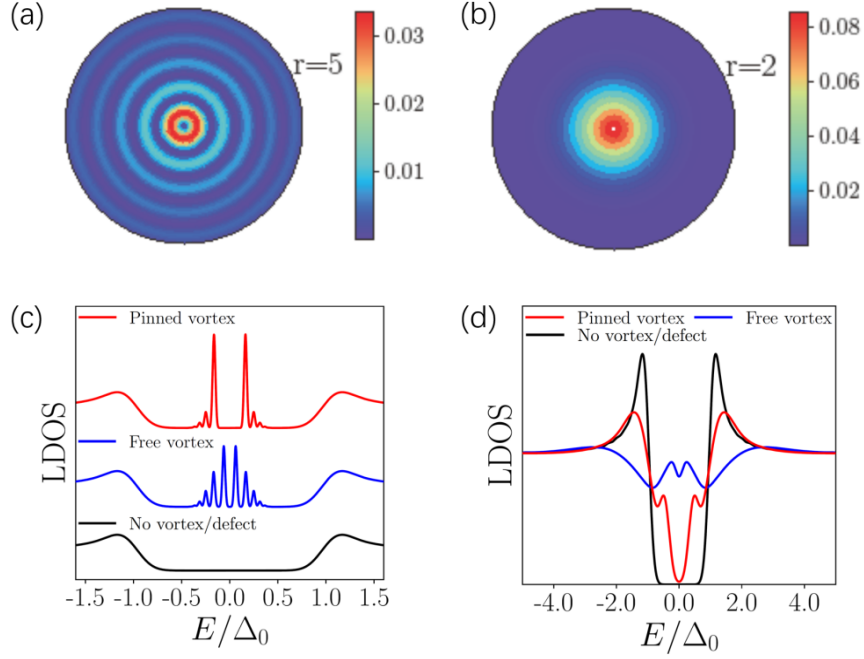

**FIG. S8.** The spatial distribution of the lowest CdGM mode in a free vortex (a) and impurity pinned vortex (b). (c) and (d) The LDOS calculated for (i) the clean SC, (ii) the free vortex (at vortex center), and (iii) the pinned vortex case (at vortex center), with taking into account the contribution from all the CdGM states, the impurity states, and the above-gap continuum. The delta function encountered in the calculations are approximated by Gaussian function with a broadening factor $b$. (c) and (d) show the results for $b = 0.002$ and $b = 0.02$, respectively. The larger broadening factor in (d) observe the discrete spectrum in (c).

## II -4. Comparison with experiments.

We now consider the more realistic case where the impurity scattering has anisotropy with nonzero $V_1$. The anisotropic scattering component $V_1$ introduces a new hybridization term in the Nambu-OAM-Bessel basis, i.e.,

$$H_1 = \sum_\sigma \int dr r V_1(r) \sum_{mn} [u^*_{m0\sigma}(r) u_{n1\sigma}(r) - v^*_{m,-1\sigma}(r) v_{n0\sigma}(r)] \gamma^\dagger_{m,-1/2} \gamma_{n1/2} + h.c., \quad (18)$$

where $V_1(\mathbf{r}) = 4V_1 \cos^2(\theta/2) e^{-(r/r_0)^2}/(\sqrt{\pi} r_0)$. Clearly, Eq. (18) introduces a direct coupling between the two lowest CdGM modes with $m = 1/2$ and $m = -1/2$. The energy spectra of the in-gap states are calculated and shown in Fig. S9(a), (b) with taking into account the nonzero $V_1$. It is found that the direct coupling between the CdGM modes with $m = 1/2$ and $m = -1/2$ always gaps out the gapless crossing point at $\Gamma_c$. The gapless point is sensitive to $V_1$, as the latter breaks the OAM conservation. The gap value with increasing $V_1$ is shown in Fig. S9(b). As shown, the gap becomes quite significant even for the case with a relatively small anisotropy, i.e., $V_1 < V_0$.

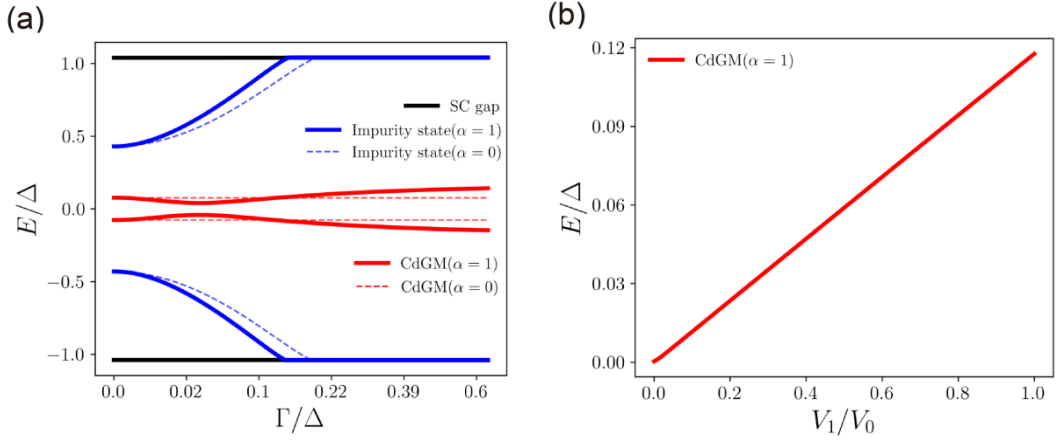

**FIG. S9.** (a) The in-gap states with anisotropic scattering. The solid and dashed data are results for $\alpha=1$ and 0 respectively. (b) The energy gap at $\Gamma_c$ caused by the anisotropy as a function of $V_1/V_0$.

Finally, to make quantitative comparison with our experiments, we now consider the impurity scattering of a finite range, i.e., $r_0 \neq 0$ in Eq. (4). $V_0(r)$ now decays with the distance $r$ from the impurity (vortex) center. In our experiments, the STM tip is gradually moved along a line-cut towards and then away from the impurity center. The distance between the tip and the impurity center is denoted by $d$. Since STM measures the local DOS at the tip site, we can use the scattering strength at $r=d$, i.e., $V_0(d)$, to simulate the local quantity measured by our experiments. With tuning the position of STM tip, $V_0(d)$ is accordingly varied, leading to a continuous evolution of the in-gap states in our calculations. The calculated results with tuning $d$ are shown in Fig. S10(a), which are compared to the experimental data marked by stars. Here, we only plot the lowest CdGM mode and the impurity state for clarity. A more complete comparison between our calculations and experiments is illustrated in Fig. S10(b) and (c). As shown, our calculation successfully accounts for the shifting behavior of the CdGM modes at quantitative level. Our calculations also predict a shifting of the impurity state to higher energy until it reaches the SC gap edge. Corresponding signatures can be found in Fig. S10(c), which is however obscured by the other in-gap CdGM states. This could be an interesting signature for further exploration.

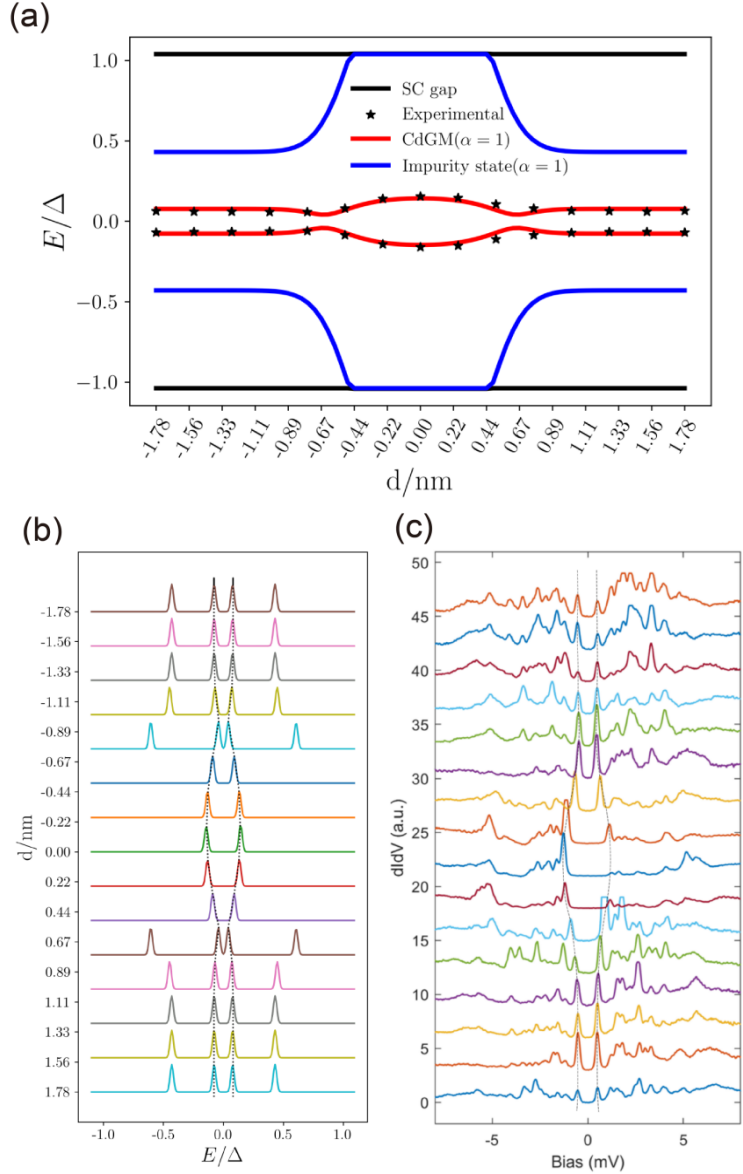

**FIG. S10.** The comparison between the numerical simulation and the STM data in experiments. The dashed curves in (b) and (c) indicate the evolution of the lowest CdGM state with moving the STM tip. The parameters used have been illustrated in the main text.

### II-5. Generalization to d-wave superconductors.

Our method and results reveal an impurity-vortex coupling mechanism explaining the "moving" of the vortex states. We now demonstrate that this mechanism is not unique for s-wave superconductors, but can be generalized to other pairing symmetries, such as the d-wave superconductors.

The d-wave SCs generally have pairing potentials with nodal points, leading to a non-fully gapped DOS. In particular, the Anderson's theorem does not apply for the d-wave cases. Both non-magnetic and magnetic impurities can act as the pair breakers. Moreover, although the vortices do not exhibit discrete bound state, they can still generate resonant-type low-energy excitations [S1], which significantly renormalize the DOS. It is intriguing to examine whether the impurity would still drive the energy shift of these vortex induced excitations.

To this end, we consider the same model in Eqs. (1)-(3), but replacing the s-wave SC order parameter by that of the $d_{x^2-y^2}$ pairing symmetry. Because of d-wave pairing symmetry, the "V"-shaped DOS is obtained at low-energies, as shown by the black dashed curve in Fig. S11(a). We further consider the vortex, whose configuration is determined by self-consistent calculations in Ref. S2, i.e.,

$$\Delta_d(\mathbf{r}) = (\Delta^{(1)}(r) + \Delta^{(2)}(r))e^{i\nu\theta} = \Delta_0(a_0 + b_0 r^3 + c_0 r^3 e^{-i4\theta})e^{i\nu\theta}, \quad (19)$$

where $\Delta^{(1)}$, $\Delta^{(2)}$ respectively represent for the isotropic and the anisotropic parts, and $a_0$, $b_0$, $c_0$ are parameters determined by model details. Then, we perform the same Bogoliubov transformation as Eq.(14)-(15) and expand the Hamiltonian by the disc Bessel functions Eq.(10). The transformed Hamiltonian has the following elements,

$$\begin{aligned}
H^{(11)}_{m_1 m_2 j_1 j_2} &= 2\pi \left( \frac{\beta^2_{m_1+\frac{1}{2},j_1}}{2mR^2} - \mu \right) \delta_{m_1 m_2} \delta_{j_1 j_2} \\
H^{(22)}_{m_1 m_2 j_1 j_2} &= -2\pi \left( \frac{\beta^2_{m_1-\frac{1}{2},j_1}}{2mR^2} - \mu \right) \delta_{m_1 m_2} \delta_{j_1 j_2} \\
H^{(12)}_{m_1 m_2 j_1 j_2} &= \Delta^{(1)}_{m_1+\frac{1}{2}, m_1-\frac{1}{2}, j_1, j_2} \delta_{m_1 m_2} + \Delta^{(2)}_{m_1+\frac{1}{2}, m_1-\frac{1}{2}+4, j_1, j_2} \delta_{m_2, m_1+4} \\
H^{(21)}_{m_1 m_2 j_1 j_2} &= \Delta^{(1)}_{m_1-\frac{1}{2}, m_1+\frac{1}{2}, j_1, j_2} \delta_{m_1 m_2} + \Delta^{(2)}_{m_1-\frac{1}{2}, m_1+\frac{1}{2}-4, j_1, j_2} \delta_{m_2, m_1-4}
\end{aligned} \quad (20)$$

where $\Delta^{(\alpha)}_{mnij} = \int r dr\, \phi^*_{mi}(r) \Delta^{(\alpha)}(r) \phi_{nj}(r)$. From Eq.(20), it is clear that the $m$ and $m \pm 4$ modes are now mixed. Thus, the orbital angular momentum $m$ is no longer a good quantum number, in contrast to the s-wave case. This well accounts for the absence of discrete CdGM states in the d-wave superconductors.

However, after transformation of $H_{\text{hyb}}$, we find that the impurity is still coupled to the two lowest vortex states characterized by $m = \pm 1/2$, which are in turn coupled to the higher CdGM states. Nevertheless, the total Hamiltonian can be numerically diagonalized. In Fig.S11(b,c), we plot the DOS (at the vortex core) with increasing the impurity coupling $V_0$. As shown by Fig.S11(b,c), with increasing the impurity scattering strength $V_0$, we find an evolution of the vortex-induced excitations. In particular, the states marked by the black arrow move towards higher energy with increasing $V_0$. Here, we note that the motion takes place for several collective modes (which form a continuum) rather than a single mode. This is because the vortex induced states with larger $m$ (i.e., $m \pm 4, m \pm 8, \ldots$) are also indirectly coupled to the impurity. Although it may be challenging to experimentally observe such an evolution in the continuum background, the evolution numerically found here clearly indicates that the impurity-vortex coupling scenario is also applicable for the d-wave superconductor.

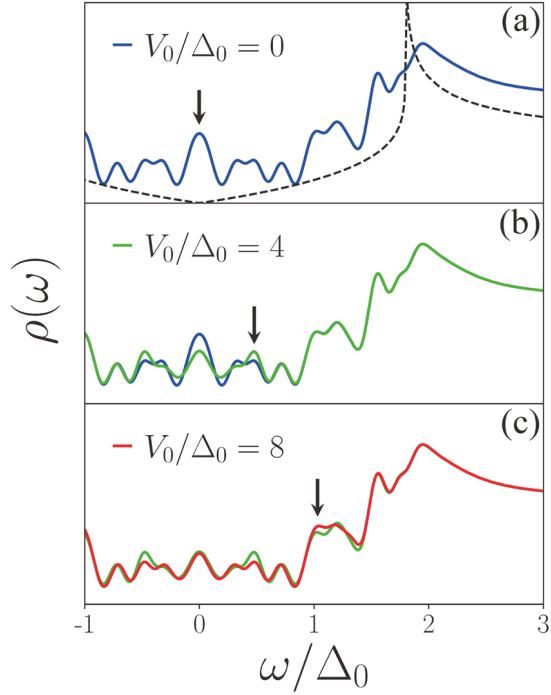

**FIG. S11.** The evolution of the vortex-induced excitations in d-wave superconductors. (a) The DOS at the vortex core in BCS d-wave superconductors, without coupling to the impurity, $V_0 = 0$. The black dashed curve shows the DOS for vortex free d-wave superconductors. (b) The DOS at the vortex core with $V_0/\Delta_0 = 4$ (green curve), compared to the that for $V_0 = 0$ (blue curve). (c) The DOS at the vortex core with $V_0/\Delta_0 = 8$ (red curve), compared to that for $V_0/\Delta_0 = 4$ (green curve). The black arrow denotes the spectral density transfer induced by the impurity coupling.